\documentclass[a4paper,11pt]{article}
\usepackage{amsfonts}

\pdfoutput=1

\usepackage{graphicx}
\usepackage{amsmath}
\usepackage{amssymb}
\usepackage{latexsym}
\usepackage[colorlinks]{hyperref}
\usepackage{color}
\usepackage{float}
\usepackage{cite}
\usepackage{makeidx}
\usepackage{colortbl}
\usepackage{csquotes}
\usepackage[squaren]{SIunits}

\usepackage[textfont={footnotesize,sf},labelfont={color=blue,bf,sf},labelsep=endash]{caption}
\usepackage[position=top,labelfont={color=blue,bf,sf}]{subfig}
\usepackage{bm}

\newcommand{\bra}{\begin{array}}
\newcommand{\era}{\end{array}}
\newcommand{\beq}{\begin{equation}}
\newcommand{\eeq}{\end{equation}}
\newcommand{\bqr}{\begin{eqnarray}}
\newcommand{\eqr}{\end{eqnarray}}

\def\BC{\bb C}
\def\_\BC{\bbi C}

\def\Tr {{\rm Tr}}

\def\( {\left(}
\def\) {\right)}
\def\no2 {{\textstyle{n\over 2}}}

\def\Tr {{\rm Tr}}

\begin{document}

\begin{titlepage}
\setcounter{page}{1}
\renewcommand{\thefootnote}{\fnsymbol{footnote}}

\begin{flushright}
%ucd-tpg 10-01\\
%arXiv:yymm.xxxx
\end{flushright}

\vspace{5mm}
\begin{center}

{\Large \bf {Band Structures of Symmetrical Graphene Superlattice \\
with Cells of three Regions}}

\vspace{5mm}
{\bf Abdellatif Kamal}$^{a}$, {\bf El Bou\^azzaoui Choubabi}$^{a}$ and
 {\bf Ahmed Jellal\footnote{\sf %ajellal@ictp.it --
a.jellal@ucd.ac.ma}}$^{a,b}$

\vspace{5mm}

{$^{a}$\em Theoretical Physics Group,  %Department of Physics,
Faculty of Sciences, Choua\"ib Doukkali University},\\
{\em PO Box 20, 24000 El Jadida, Morocco}

%{$^b$\em MATIC, FPK, Hassan 1 University, Khouribga, Morocco}

{$^b$\em Saudi Center for Theoretical Physics, Dhahran, Saudi Arabia}

%{$^{d}$\em Max Planck Institute for the Physics of Complex Systems,\\
%N\"othnitzer Str. 38, 01187 Dresden} %\\[1em]

\vspace{30mm}

\begin{abstract}
  We study the electronic band structures of massless Dirac fermions in
  symmetrical graphene superlattice
  with cells of three regions.
  %composed of three regions according to the applied potential. 
  Using the transfer matrix method, we explicitly determine the dispersion relation in terms of different physical parameters. We numerically analyze such relation and show that there exist three zones: bound, unbound and forbidden states. In the central zone of the band structures, we determine and enumerate the vertical Dirac points, %$(k_y=0)$, 
  opening gaps and additional Dirac points.
  %only in minibands. 
  Finally, we inspect the potential effect on minibands, the anisotropy of group velocity and the energy bands contours near Dirac points. We also discuss the evolution of  gap edges and {cutoff region} near the vertical Dirac points.

\vspace{3cm}



\end{abstract}
\end{center}
\end{titlepage}

%======================================================================
\section{Introduction}
%======================================================================

Graphene \cite{Novoselov666,RN2} is one of the most  interesting two-dimensional system realized in modern physics
right now. This is particularly due  
the fact that the
physics of low-energy carriers in graphene is governed by
a Dirac-like Hamiltonian and carriers are massless fermions of band structures  
without energy gap
\cite{RevModPhys.83.851,PhysRevB.65.245420,PhysRevLett.95.146801}. 
As a consequence of this relativistic-like behavior particles could tunnel through 
very high barriers
in contrast to the conventional tunneling of non-relativistic particles, an effect known in relativistic
field theory as Klein tunneling. This effect has already been observed experimentally 
\cite{2}
in graphene systems. There are various ways for creating barrier structures in graphene \cite{3,4}. For
instance, it can be done by applying a gate voltage, cutting the graphene sheet into finite width to
create a nanoribbons, using doping or through the creation of a magnetic barrier.

To understand the graphene basic physical properties, 
 numerous works for single layer graphene superlattices (SLGSLs)
have been appeared studying the band structure theory for various kinds of periodic potentials \cite{CastroNeto2009}.
Increasing interest in SLGSLs results from the prediction of possibly engineering the system band
structure by the periodic potential, which opens different
ways to fabricate graphene-based electronic devices. Experimentally,
the profile of periodic potentials (graphene-based superlattices) can be realized by interaction with a substrate \cite{PhysRevB.76.075429,PhysRevLett.100.056807} or controlled by adatome deposition \cite{Jannik2008}. Theoretically, 
it can be generated by application
 of the electrostatic potential \cite{Park2008,ParkCheol2008,ParkCheol2009,Brey2009,Barbier2010,Pham2010} and magnetic barriers \cite{Ghosh2009,Ramezani2010,Snyman2009,DellAnna2011,VQui2012}. 
 In fact, graphene-based superlattices formed by modulated potentials were studied 
 using different approaches. 
 The corresponding 
 electronic band structure of electric and magnetic SLGSLs showed emergence of additional Dirac points  and a strong anisotropic renormalization of the carrier group velocity 
 \cite{Park2008,ParkCheol2009,Brey2009,Barbier2010,Pham2010}. 
It is also showed that SLGSLs have important properties for electronic applications such as an  excellent electric conductivity and possibility to control the electronic structure externally. 

The purpose of the present work is to study the electronic band structures of single layer graphene superlattice
formed by identical cells of three regions (SLGSL-3R) using transfer matrix approach. This generalizes that of 
a graphene-based superlattice formed by a Kronig-Penny periodic potential  \cite{Pham2010} where
a new region is  introduced between the barrier and well, called central region, according to the potential profile in Figure 
\ref{Figmodel:SubFigA}. 
After matching different regions at interfaces, 
we explicitly determine the dispersion relation in terms of different physical parameters. These allow
us to 
%as well as 
locate the emergence of the horizontal contact points (HCPs) and vertical contact points (VCPs).
Around such points, we 
expand the dispersion relation to easily evaluate the group velocity components
and therefore show that the contact points are really the emerged  Dirac points in SLGSL-3R.
Analyzing numerically the dispersion relation, 
we show that there exist three zones: bound, unbound and forbidden states. Subsequently, we determine and enumerate VDPs $(k_y=0)$ and opening gaps  in the central zone of the band structures
as well as  Dirac points in minibands. 
Finally, we study the potential effect on minibands, the anisotropy of group velocity and the energy bands contours near Dirac points as well as discuss the evolution of edge gaps and {cutoff region} near VDPs.

The present paper is organized as follows. In section \ref{Sec:Model}, we  introduce the theoretical model 
describing SLGSL-3R and determine the energy spectrum for one cell composed of three regions. This will be used to
explicitly obtain the dispersion relation for the whole SLGSL-3R using transfer matrix method.
We numerically analyze the electronic band structures and therefore underline
the main features exhibited by our system in section \ref{Sec:Results}.
We conclude our work and summarize the obtained results 
in section \ref{Sec:Conclusion}.
Finally, our paper is closed by an appendix dealing with the contact points location in the center and edges of the first Brillouin zone. With these we will be able to 
derive an approximate dispersion relation and therefore obtain the 
corresponding normalized group velocities.

%======================================================================
\section{Theoretical model}\label{Sec:Model}
\begin{figure}[!ht]
\centering
\subfloat[]{
    \includegraphics[scale=0.8]{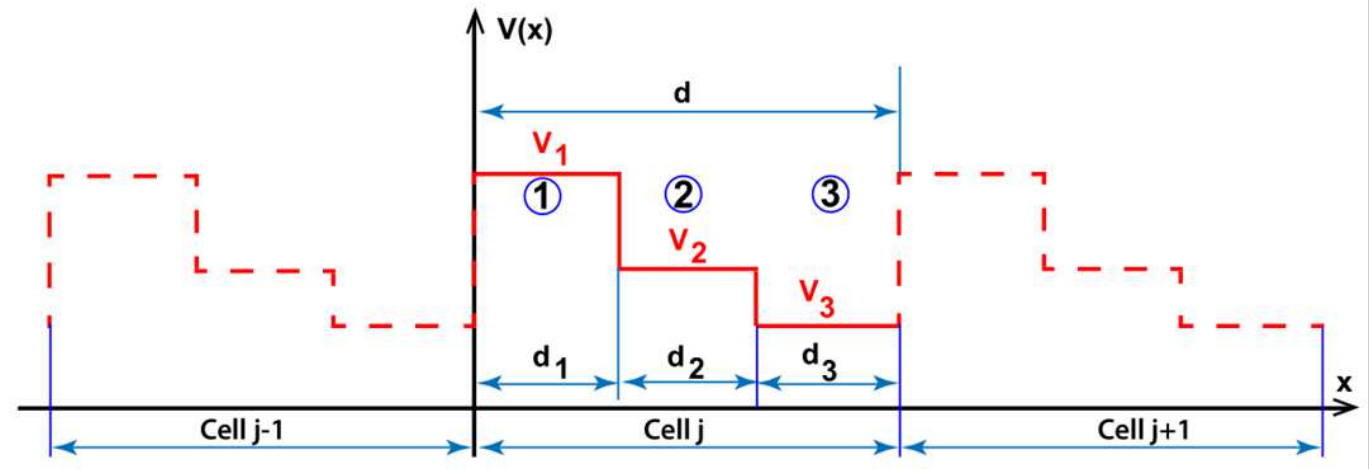}
    \label{Figmodel:SubFigA}
}\\
\subfloat[]{
    \includegraphics[scale=0.8]{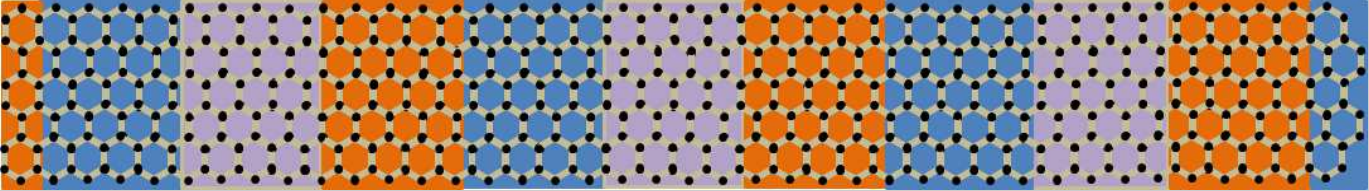}
    \label{Figmodel:SubFigB}
}
  \caption{(Color online) \protect\subref{Figmodel:SubFigA}: The superlattice potential $V(x)$ composed of three amplitudes $(V_1, V_2, V_3)$
  and each one is associated to a given region $i$.
  %regions, which 
  It is growing along the $x$-direction with the period $d=d_1+d_2+d_3$ and $d_i$ is the width of region $i$. 
  %The potential $V(x)$ is applied to graphene with amplitude $V_i$ in region $i$. 
  \protect\subref{Figmodel:SubFigB}: Showing a single layer graphene under $V(x)$ deposed on silicone dioxide $(SiO_2)$ substrate where each color represented on the substrate corresponds to a given region.}\label{FigModel}
\end{figure}

%======================================================================
To neglect the edge effects, we consider an infinitely large single layer graphene deposited on a substrate of silicon dioxide $(SiO_2)$ (Figure \ref{Figmodel:SubFigB}) and subjected to a very long periodic potential (Figure \ref{Figmodel:SubFigA}) along the $x$-direction. It contains $n$ elementary cell labeled by $j$ $(j=0, \cdots,n-1)$ and each one is composed by a juxtaposition of three single square barriers with different height $(V_1, V_2, V_3)$ and width $(d_1,d_2 ,d_3)$, with $d=d_1+d_2 +d_3$ is the width of the entire cell.

\noindent
The $j$-th elementary cell has  two internal junctions in the positions $\left(x_2=jd+d_1, x_3=jd+d_1+d_2\right)$ and two extreme junctions in the positions $\left(x_1=jd, x_4=\left(j+1\right)d\right)$. According to Figure \ref{Figmodel:SubFigA}, for $j$-th elementary cell the potential is given by
\begin{equation}%\label{V}
V^j(x)=
    \left\{
        \begin{array}{cccc}
            V_1&\text{if}   & jd < x < d_1+jd\\
            V_2&\text{if}   & d_1+jd < x < d_1+d_2+jd\\
            V_3&\text{if}   & d_1+d_2+jd < x < (1+j)d\\
        \end{array}
    \right.
\end{equation}
Our graphene superlattice consists of $n$ elementary cells, which are interposed between the \emph{input} and \emph{output} regions described by the free  Hamiltonian
\begin{equation}\label{Hinout}
  H_{in/out}=v_F\bm{\sigma}\cdot\bm{p}
\end{equation}
while, each region $i$ of $j$-th elementary cell  (Figure \ref{Figmodel:SubFigA}) has the Hamiltonian
\begin{equation}%\label{002}
H_{i}^{j}=v_F\bm{\sigma}\cdot\bm{p}+V_{i}^{j}\mathbb{I}
\end{equation}
where $\bm{p}=(p_x,p_y)$ is the momentum operator, $v_F\approx 10^6 \meter\per\second$ the Fermi velocity, $\bm{\sigma}=(\sigma_x,\sigma_y)$ the Pauli matrices, $\mathbb{I}$ the $2\times 2$ unit matrix and the index $i=1, 2, 3$.

For a given cell $j$, we can solve the eigenvalue equation
\begin{equation}%\label{03}
\left(%
\begin{array}{cc}
  V_{i}-E_{i} & v_F(p_x-\textbf{\emph{i}} p_y) \\
 v_F( p_x+\textbf{\emph{i}} p_y) & V_{i} -E_{i}\\
\end{array}%
\right)\left(%
\begin{array}{c}
  \varphi_i^A \\
   \varphi_i^B \\
\end{array}%
\right)=0
\end{equation}
with the eigenespinors $\psi_{i}(x,y) =\left(%
\begin{array}{cc}
  \varphi_i^A & \varphi_i^B \\
\end{array}%
\right)^{T}$ and $\varphi_i^{A/B}$ are smooth enveloping functions for two triangular sublattices ($A$, $B$) in graphene, which take the forms $\varphi_i^{A/B}(x) e^{\textbf{\emph{i}}k_y y}$ due to the translation invariance in the $y$-direction. It is convenient to introduce the dimensionless quantities $\mathbb{V}_i=V_i/E_F$, $\varepsilon_i=E_i/E_F$ with $E_F=\hbar v_F/d$ and therefore getting two dependent equations
\begin{eqnarray}\label{04A}
  &&  -\left(\varepsilon_i-\mathbb{V}_i\right)\varphi_i^A +\frac{d}{\hbar}(p_x-\textbf{\emph{i}} \hbar k_y) \varphi_i^B=0
\\
&&
\label{04B}
    -(\varepsilon_i-\mathbb{V}_i)\varphi_i^B +\frac{d}{\hbar}(p_x+ \textbf{\emph{i}}\hbar k_y) \varphi_i^A=0
\end{eqnarray}
giving rise the second order differential equations
\begin{equation}%\label{05}
  \frac{d^2}{\hbar^2}(p_x^{2}+\hbar^{2}k_y^{2}) \varphi_i^{A,B}=(\varepsilon_i-\mathbb{V}_i)^{2} \varphi_i^{A,B}.
\end{equation}
For the first spinor component $\varphi_i^A(x)$, the solution reads as
\begin{equation}%\label{pml}
  \varphi_i^A=\alpha_i e^{ \textbf{\emph{i}} k_{i} x}+\beta_i e^{-\textbf{\emph{i}} k_{i} x}
\end{equation}
where the wave-vector along the direction of propagation in each region $i$ is giving by
\begin{equation}\label{h0}
  k_i=\frac{1}{d}\sqrt{(\varepsilon-\mathbb{V}_i)^2-(k_yd)^2}
\end{equation}
and $ \alpha_i$, $\beta_i$ are the amplitude of positive and negative propagation wave-functions inside the region $i$, respectively.
Using \eqref{04A} to obtain the second spinor component
\begin{equation}%\label{07}
\varphi_i^B = \alpha_i s_i e^{ \textbf{\emph{i}} \theta_i} e^{\textbf{\emph{i}} k_i x}-
\beta_i s_i e^{ -\textbf{\emph{i}} \theta_i} e^{- \textbf{\emph{i}} k_i x}
\end{equation}
where we have set $\theta_{i} = \arctan \left(\frac{k_y}{k_{i}}\right)$. Finally, the general solution takes the form
\begin{equation}\label{psixy}
\psi_{i}(x,y)=\psi_i(x)e^{\textbf{\emph{i}}k_y y}
\end{equation}
and the function along $x$-direction can be written as
\begin{equation}\label{12}
\psi_{i}(x)=w_i(x)D_i
\end{equation}
such that the involved quantities are
\begin{equation}\label{equ13}
  w_{i}(x)=\left(%
\begin{array}{cc}
  e^{ \textbf{\emph{i}} k_{i} x} & e^{ -\textbf{\emph{i}} k_{i} x} \\
  s_{i} e^{ \textbf{\emph{i}} \theta_{i}} e^{ \textbf{\emph{i}} k_{i} x}  & -s_{i} e^{ -\textbf{\emph{i}} \theta_{i}} e^{- \textbf{\emph{i}} k_{i} x} \\
\end{array}
\right),\qquad D_i=\left(%
\begin{array}{c}
  \alpha_i \\
  \beta_i \\
\end{array}
\right)
\end{equation}
with $s_i=\text{sign}(\varepsilon-\mathbb{V}_i)$. The solution of the energy spectrum  of the Hamiltonian \eqref{Hinout} can easily be derived from that obtained in \eqref{h0} and \eqref{psixy} as a particular case. This is
\begin{equation}%\label{inout}
  \psi_{in/out}(x,y)=w_0(x)e^{\textbf{\emph{i}}k_y y}D_{in/out}.
\end{equation}
Note that, our graphene superlattice is adiabatic and does not provide an exchange of energy with the external environment, which leaves the Fermi energy of the Dirac fermions preserved $E_{i}=E$ $(\varepsilon_i=\varepsilon)$ during propagation via all the elementary cells of the superlattice.

Recall that the obtained dispersion relation \eqref{h0} is for one region $i$, now it is natural to ask about that of the whole charge carriers in single layer graphene superlattice with three regions (SLGSL-3R). In getting such total dispersion relation, we use transfer matrix technique and apply the Bloch theorem. Indeed, first we determine the transfer matrix by using the continuity of the eigenspinors at different interfaces, which gives, for a unit cell ($n=1$, $j=0$), the relations
\begin{eqnarray}
  &&w_0(0)D_{in} = w_1(0 )D_1 \\
  &&w_1(d_1)D_1 = w_2(d_1)D_2 \\
  &&w_2(d_1+d_2)D_2 = w_3(d_1+d_2)D_3 \\
  &&w_3(d)D_3 = w_0(d)D_{out}.
\end{eqnarray}
Now connecting the amplitudes of propagation of the \emph{input} $D_{in}=D_0$ with those of the \emph{output} $D_{out}=D_{d}$ of the unit cell to obtain
\begin{equation}
     D_{in}= \mathcal{T}_{1}D_{out}
\end{equation}
where the transfer matrix takes the form
\begin{equation}%\label{T0}
  \mathcal{T}_1=w_0^{-1}(0)\ \Omega\ w_0(d)
\end{equation}
and $\Omega$ is given by
\begin{eqnarray}\label{omega}
\Omega&=&w_1(0)w_1^{-1}(d_1)w_2(d_1)w_2^{-1}(d_1+d_2)w_3(d_1+d_2)w_3^{-1}(d)\\
&=&w_1(jd)w_1^{-1}(d_1+jd)w_2(d_1+jd)w_2^{-1}(d_1+d_2+jd)w_3(d_1+d_2+jd)w_3^{-1}((1+j)d).
\end{eqnarray}
Using the same iterative method of small to very large $n$, we find the general relation connecting \emph{input} and \emph{output} amplitudes
\begin{equation}
     D_{in}= \mathcal{T}_{n}D_{out}
\end{equation}
where now $D_{out}=D_{nd}$ and the transfer matrix associated with $n$ identical unit cells reads as
\begin{equation}%\label{17}
 \mathcal{T}_{n}=w_{0}^{-1}(0)\ \Omega^{n}\ w_{0}(nd).
\end{equation}
Calculating the determinant and trace of 
\eqref{omega} to end up with the results
\begin{eqnarray}\label{bahraoui}
\label{18}
 \det(\Omega) &=& 1\\
  \Tr(\Omega) &=& 2 \left[\cos(k_1d_1)\cos(k_2d_2)\cos(k_3d_3)
    +G_{12}\sin(k_1d_1)\sin(k_2d_2)\cos(k_3d_3)\right.\\
    &&+\left.G_{13}\sin(k_1d_1)\sin(k_3d_3)\cos(k_2d_2)
    +G_{23}\sin(k_2d_2)\sin(k_3d_3)\cos(k_1d_1)\right]\nonumber\label{21}
\end{eqnarray}
where we have set the quantity
\begin{equation}%\label{Gij}
      G_{ij}=\frac{(\mathbb{V}_i-\mathbb{V}_j)^2-(k_i^2+k_j^2)d^2}{2k_ik_jd^2},\qquad k_{i,j}\neq 0.
\end{equation}
Recall that SLGSL-3R is periodic, then one has to use the Bloch theorem and write the spinor \eqref{12} along $x$-direction as
\begin{equation}%\label{28}
    ^B\psi_i(x) = \psi_i(x) e^{\textbf{\emph{i}}k_x x}
\end{equation}
where $k_x$ is the Bloch wave vector. Therefore, the periodic condition at the extremity of the $j$-th cell imposes
\begin{equation}
  ^B\psi_1(x_1^+)=\; ^B\psi_3(x_4^-)\label{Q22}
\end{equation}
and now the continuity conditions at interfaces give
\begin{eqnarray}
       && ^B\psi_1(x_2^-)=\; ^B\psi_2(x_2^+)\label{Q33}\\
       && ^B\psi_2(x_3^-)=\; ^B\psi_3(x_3^+).\label{Q44}
\end{eqnarray}
Because of the system periodicity, we can shift a potential cell to the origin $x=0$, which corresponds to take $j=0$. In this case \eqref{Q22}, \eqref{Q33} and \eqref{Q44} become
\begin{eqnarray}%\label{30}
 &&w_1(0)D_1        = w_3(d)D_3 e^{ \textbf{\emph{i}} k_x d}\\
 &&w_1(d_1)D_1      = w_2(d_1)D_2\\
 &&w_2(d_1+d_2)D_2  = w_3 (d_1+d_2)D_3.
\end{eqnarray}
They can be used  to express $D_3$ in terms of $D_1$ as
\begin{eqnarray}
  D_3 &=& w_3^{-1}(d)w_1(0)D_1 e^{- \textbf{\emph{i}} k_x d} \label{311} \\
   &=& w_3^{-1}(d_1+d_2)w_2(d_1+d_2)w_2^{-1}(d_1)w_1(d_1)D_1 \label{312}.
\end{eqnarray}
Now using \eqref{equ13} together with \eqref{311} and \eqref{312} to obtain the relation
\begin{equation}\label{32}
   M\left(
\begin{array}{c}
 \alpha _1 \\
 \beta _1 \\
\end{array}
\right)=0
\end{equation}
where the matrix $M$ is giving by
\begin{equation}
  M=w_3^{-1}(d)w_1(0)e^{- \textbf{\emph{i}} k_x d}-w_3^{-1}(d_1+d_2)w_2(d_1+d_2)w_2^{-1}(d_1)w_1(d_1).
\end{equation}
The non-trivial solution of  \eqref{32} corresponds to $\det(M)=0$. Doing this, to end up with the dispersion relation for SLGSL-3R
\begin{eqnarray}\label{33}
 \cos(k_x d)&=&\cos(k_1d_1)\cos(k_2d_2)\cos(k_3d_3)
    +G_{12}\sin(k_1d_1)\sin(k_2d_2)\cos(k_3d_3)\nonumber\\
    &&+G_{13}\sin(k_1d_1)\sin(k_3d_3)\cos(k_2d_2)
    +G_{23}\sin(k_2d_2)\sin(k_3d_3)\cos(k_1d_1)
\end{eqnarray}
which can be written using \eqref{bahraoui}  as
\begin{equation}%\label{34}
  \cos(k_x d)=\frac{1}{2}\Tr(\Omega).
\end{equation}
Note that, an interesting limiting case can be derived from our results \eqref{33} by requiring $d_2=0$, which gives the dispersion relation of single layer graphene superlattice with two regions of potential (SLGSL-2R) \cite{Pham2010}
\begin{equation}\label{limit33}
%\begin{array}{c}
 \cos(k_x d)=\cos(k_1d_1)\cos(k_3d_3)+\frac{(\mathbb{V}_1-
 \mathbb{V}_3)^2-(k_1^2+k_3^2)d^3}{2k_1k_3d^3}\sin(k_1d_1)\sin(k_3d_3).
%\end{array}
\end{equation}
In the forthcoming analysis, we focus on the numerical study of the dispersion relation \eqref{33} in order to extract more information about our SLGSL-3R. Also different comparisons with respect to \eqref{limit33} and discussions will be reported.

%======================================================================
\section{Results and discussions}\label{Sec:Results}
%======================================================================

We will numerically analyze the dispersion relation \eqref{33} in terms of the involved parameters ($q$, $\mathbb{V}$, $d$) characterizing the applied potential. To describe the symmetric effect of such potential, we introduce the distance $q_i\!=\!d_i/d$ with $i=1, 2, 3$ and $0\!\leq\! q_i\!\leq\! 1$, giving rise $q=\left\{q_1,q_2,q_3\right\}$. For simplicity, we choose the configuration of symmetrical SLGSL-3R (SSLGSL-3R) $\left(q=\left\{\frac{1-q_2}{2},q_2,\frac{1-q_2}{2}\right\}, \mathbb{V}_1=-\mathbb{V}_3\equiv \mathbb{V}, \mathbb{V}_2\equiv0\right)$  with $d=10~\nano\meter$ in order to carry out our calculations.

%======================================================================
\subsection{Electronic band structures}
%======================================================================

\begin{figure}[!ht]
\centering
\subfloat[]{
    \includegraphics[scale=0.5]{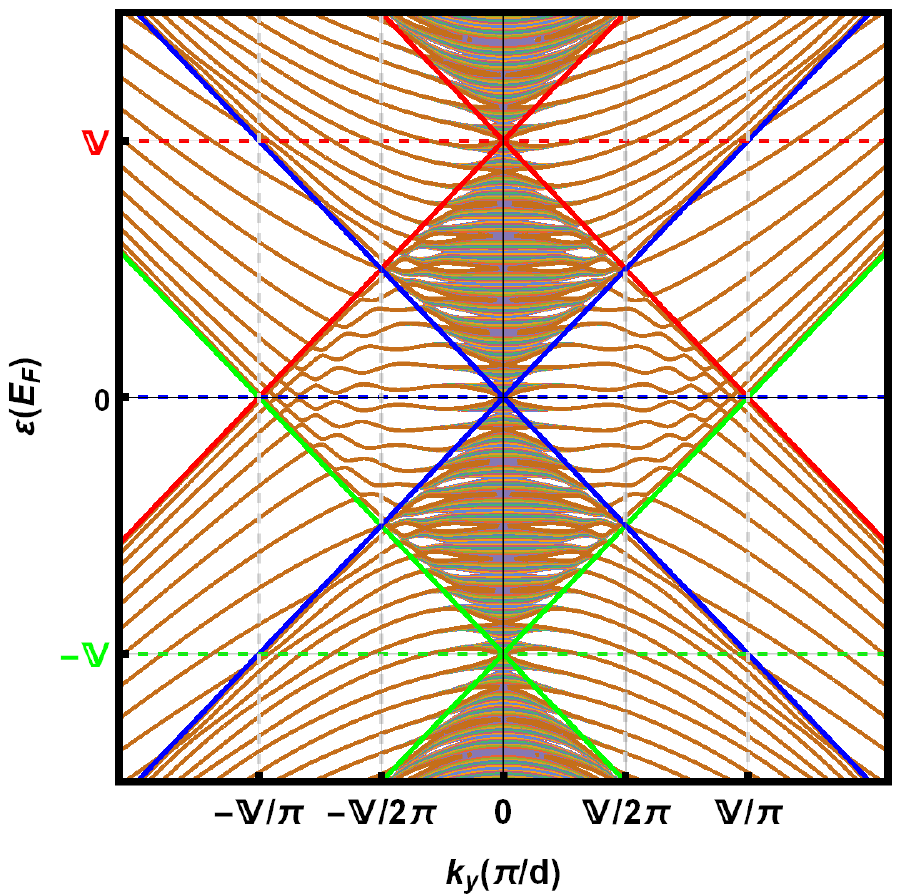}
    \label{Fig2:SubFigA}
}\hspace{20pt}
\subfloat[]{
    \includegraphics[scale=0.5]{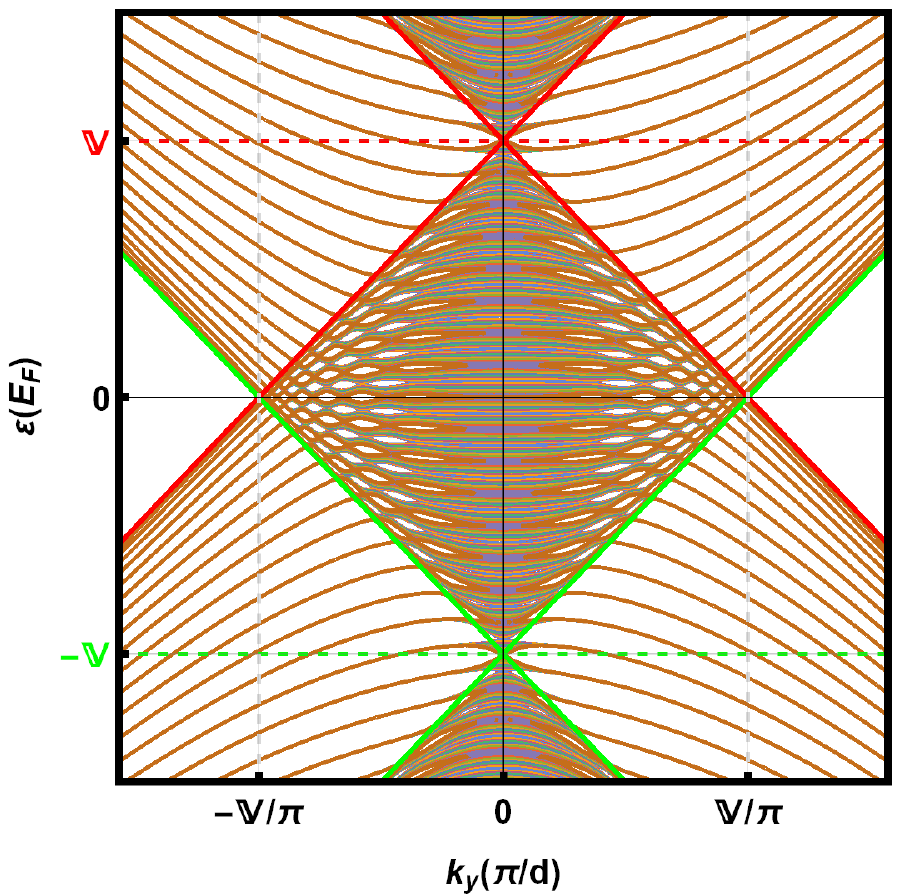}
    \label{Fig2:SubFigE}
}\\
\subfloat[]{
    \includegraphics[scale=0.5]{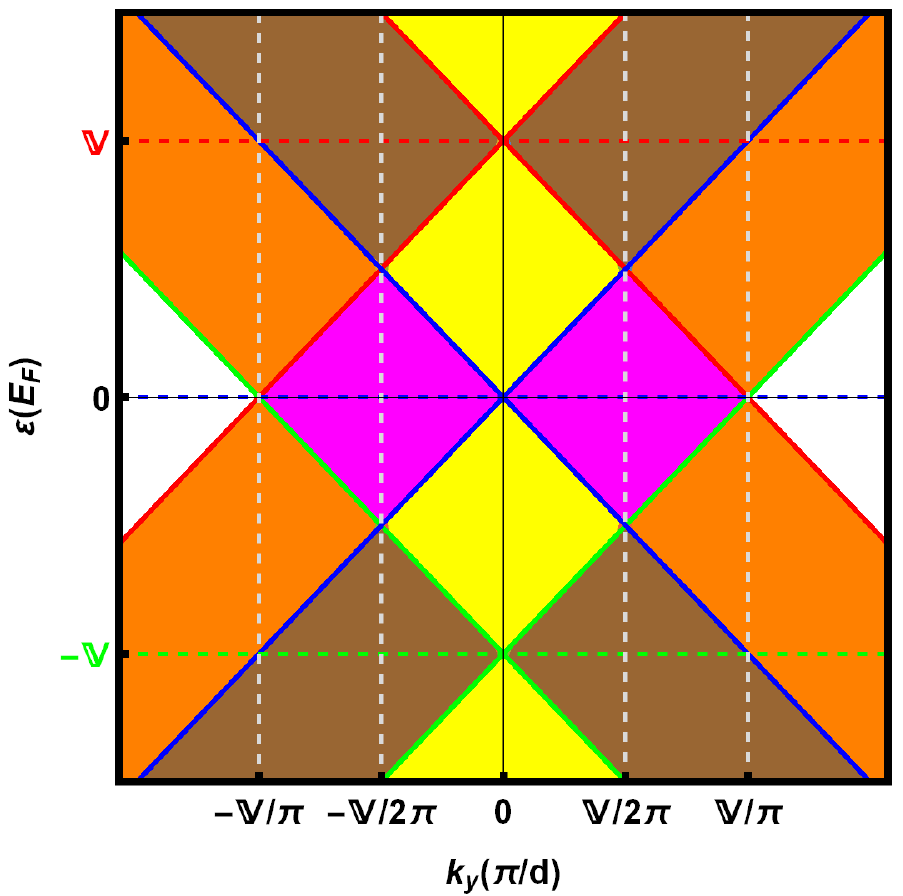}
    \label{Fig2:SubFigB}
}\hspace{20pt}
\subfloat[]{
    \includegraphics[scale=0.5]{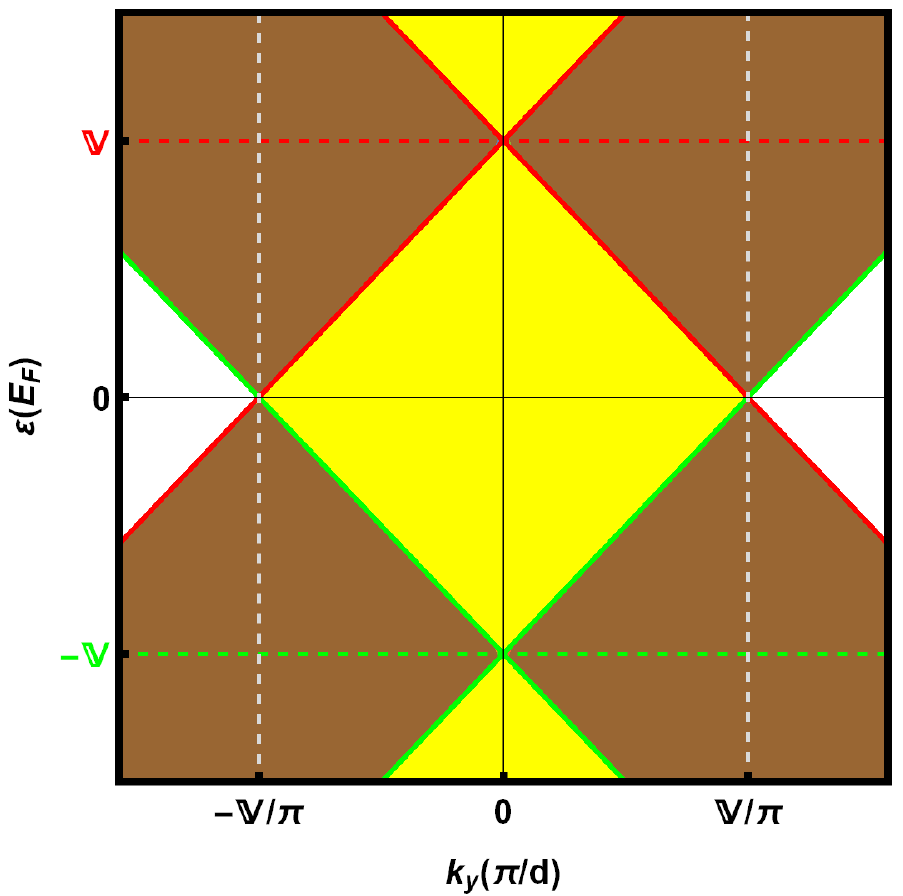}
    \label{Fig2:SubFigF}
}
\caption{(Color online) General behavior of electronic band structures for  \protect\subref{Fig2:SubFigA}: SSLGSL-3R with $q_2=\frac{1}{3}$ and \protect\subref{Fig2:SubFigE}: SSLGSL-2R  $\left(q_2=0\right)$. Different zones of the electronic band structure for \protect\subref{Fig2:SubFigB}: SSLGSL-3R and \protect\subref{Fig2:SubFigF} SSLGSL-2R.
}
\label{StructAband1}
\end{figure}
Figure \ref{StructAband1} illustrates the electronic band structures for SSLGSL-3R versus the wave vector $k_y$ for the case $q_2=\frac{1}{3}$. The three potentials ($\mathbb{V}$, $\mathbb{V}_{2}=0$, $-\mathbb{V}$) give three cones colored by red, blue and green (Figures \ref{Fig2:SubFigA} and \ref{Fig2:SubFigB}), respectively. The combination of these cones delivers five different zones of yellow, magenta, brown, orange and white colors (Figure \ref{Fig2:SubFigB}). The zone of unbound states (yellow) consists of two lozenges and two half-lozenges, one and half-lozenge are at positive energy and one and half-lozenge at negative energy. The zone of non-parabolic bound states (magenta) contains opening gaps. The central zone (CZ) is formed by two yellow and two magenta lozenges. Brown zones contain serried parabolic  bound states, which come directly from the yellow lozenges of the unbound states. The orange zones contain less serried bound states that come from brown or magenta zones, whereas the white zone is forbidden. Note that, there are horizontal and vertical mirror symmetries, i.e. $\tau \varepsilon (\kappa k_y)= \text{const}$ with $\tau,\kappa=\pm 1$. Figures \ref{Fig2:SubFigE} and \ref{Fig2:SubFigF} show the electronic band structure behavior of symmetrical SLGSL-2R (SSLGSL-2R) where the superposition of the two cones red and green generate three distinct zones illustrated by the yellow, brown and white colors. These zones contain unbound, bound and forbidden states, respectively. The band structure does not contain non-parabolic bound states.

\begin{figure}[!ht]
\centering
\subfloat[$l=0$]{
    \includegraphics[scale=0.4]{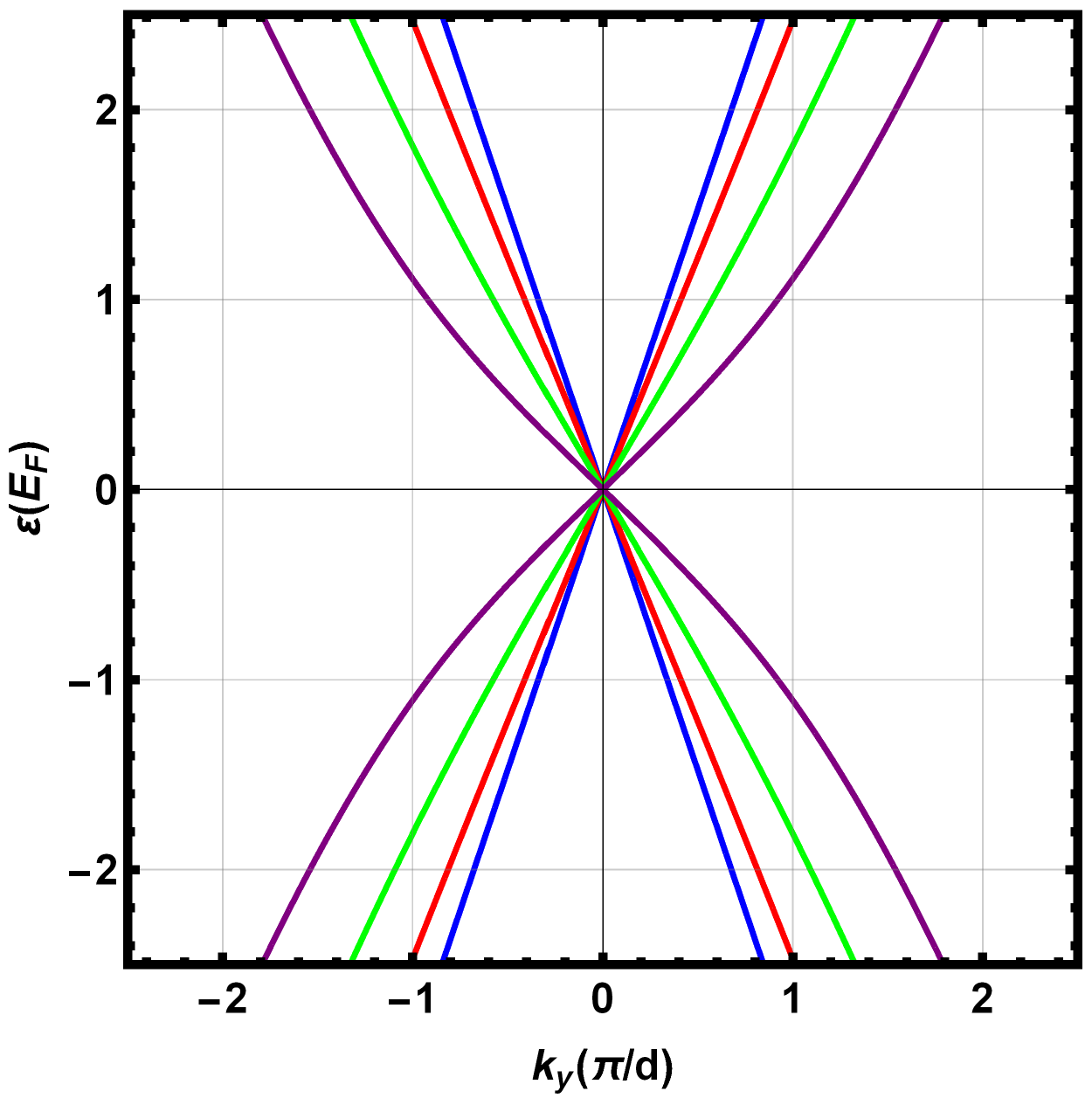}
    \label{FigCh33:SubFigD}
}
\subfloat[$l=1$]{
    \includegraphics[scale=0.4]{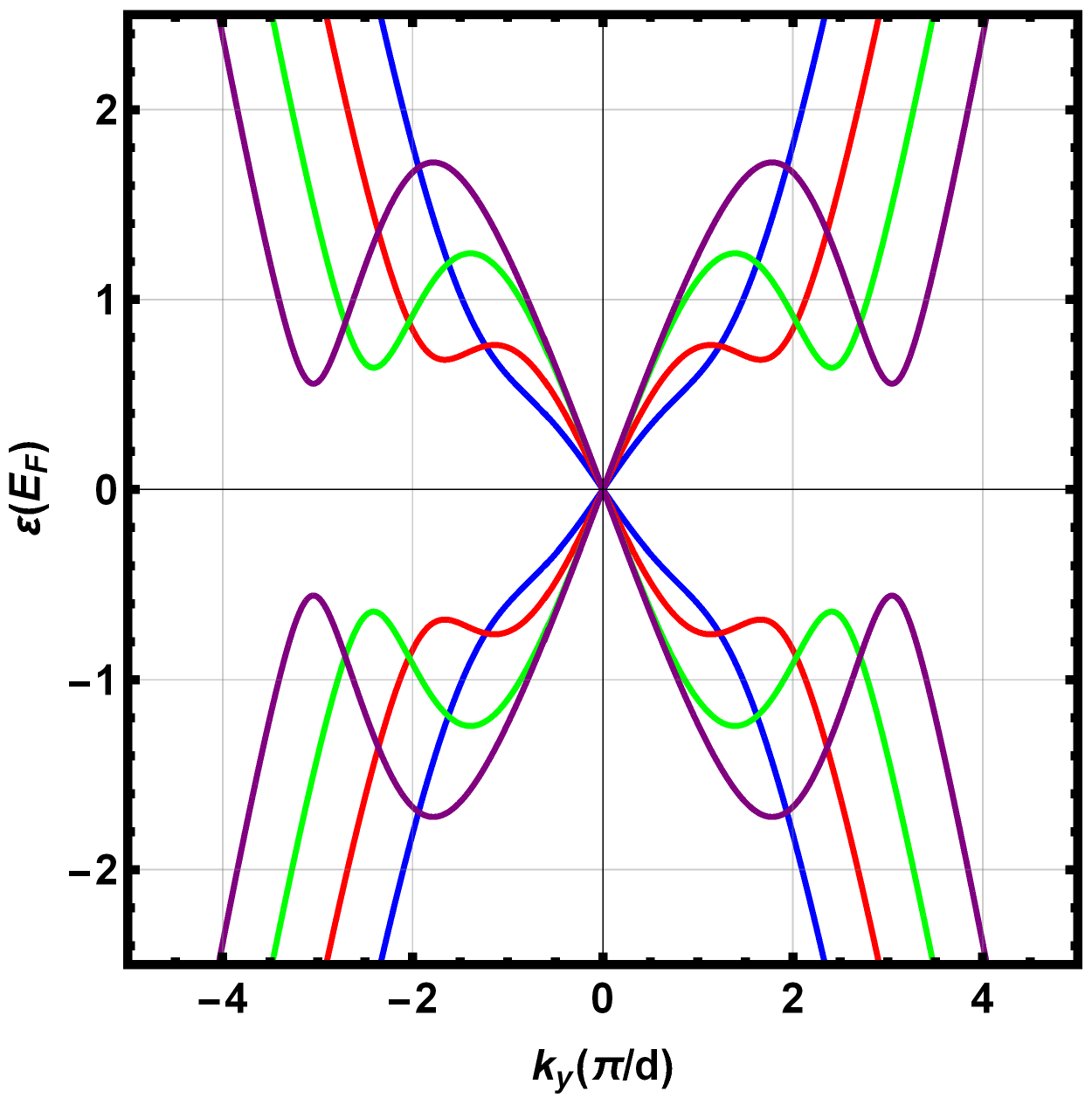}
    \label{FigCh33:SubFigE}
}
\subfloat[$l=2$]{
    \includegraphics[scale=0.4]{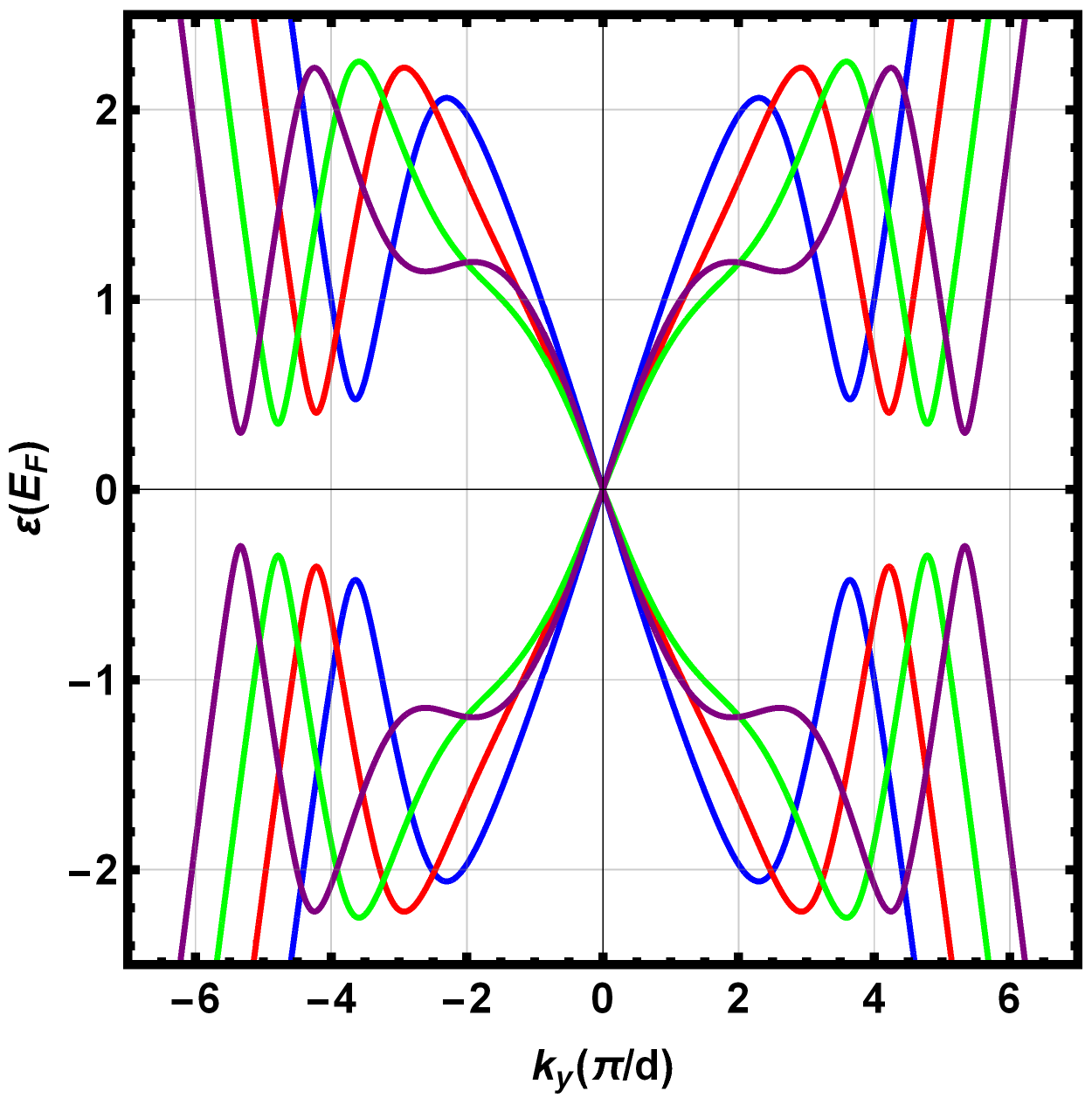}
    \label{FigCh33:SubFigF}
}\\
\subfloat[$l=0$]{
    \includegraphics[scale=0.4]{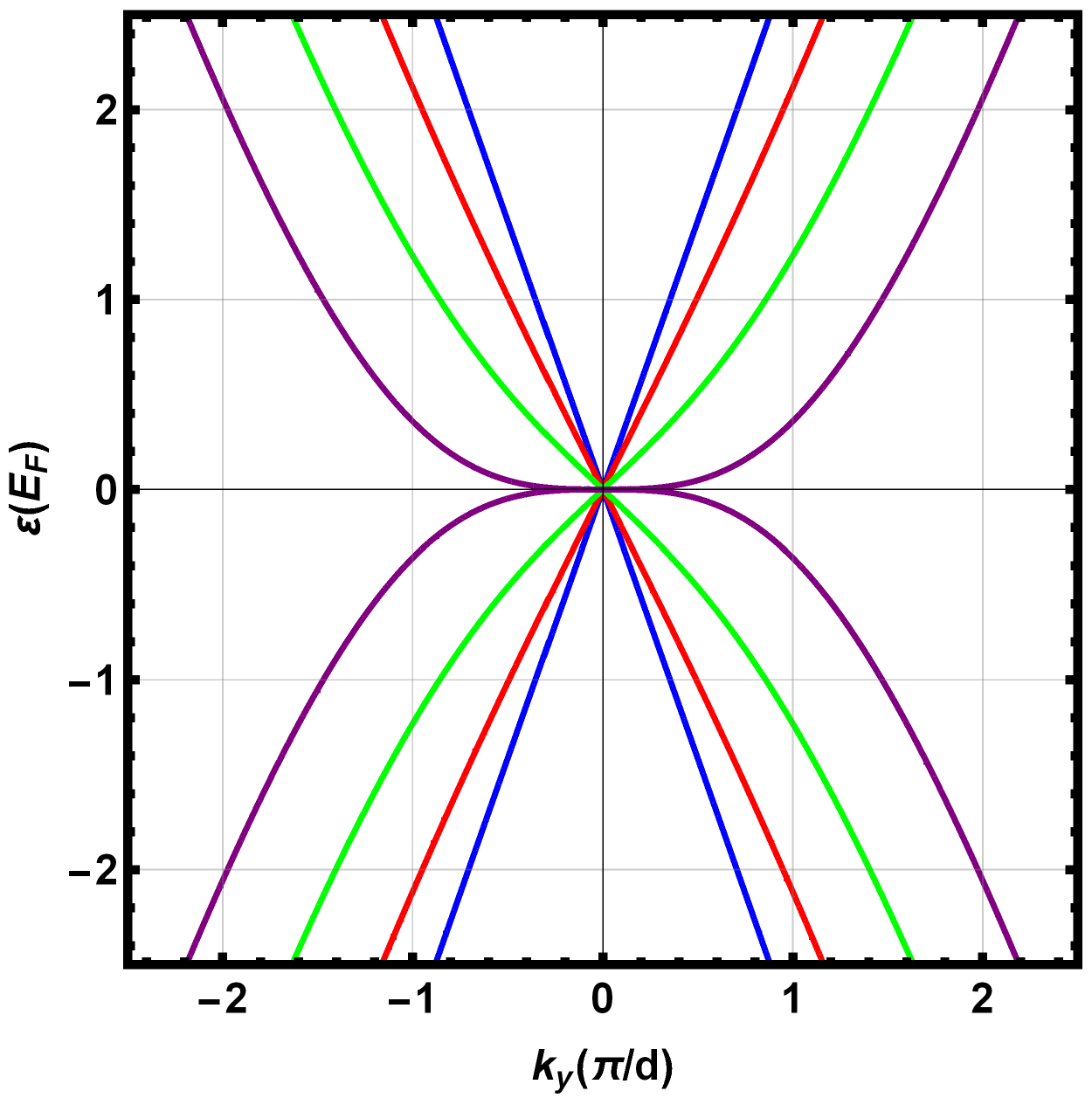}
    \label{FigCh33:SubFigA}
}
\subfloat[$l=1$]{
    \includegraphics[scale=0.4]{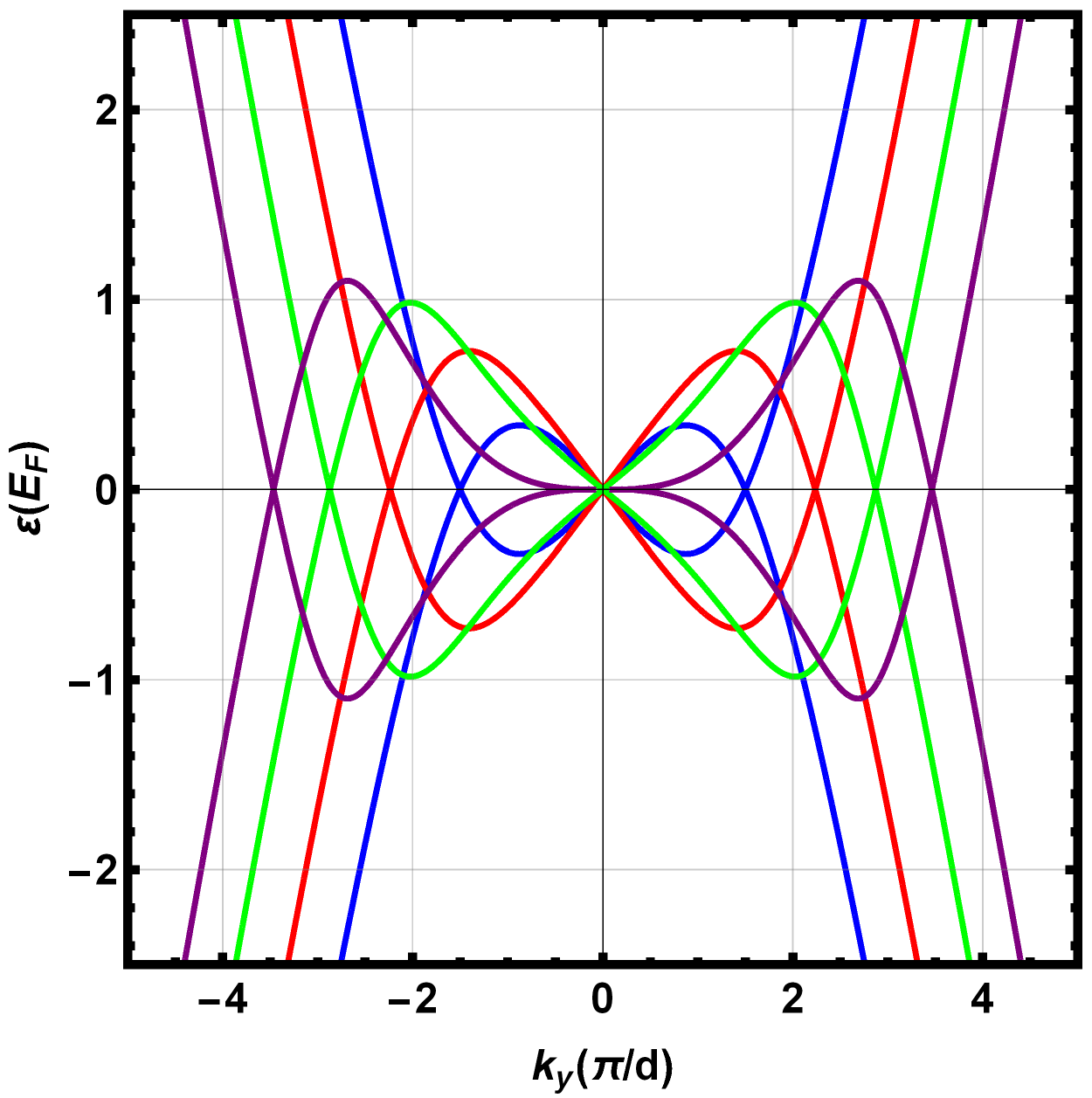}
    \label{FigCh33:SubFigB}
}
\subfloat[$l=2$]{
    \includegraphics[scale=0.4]{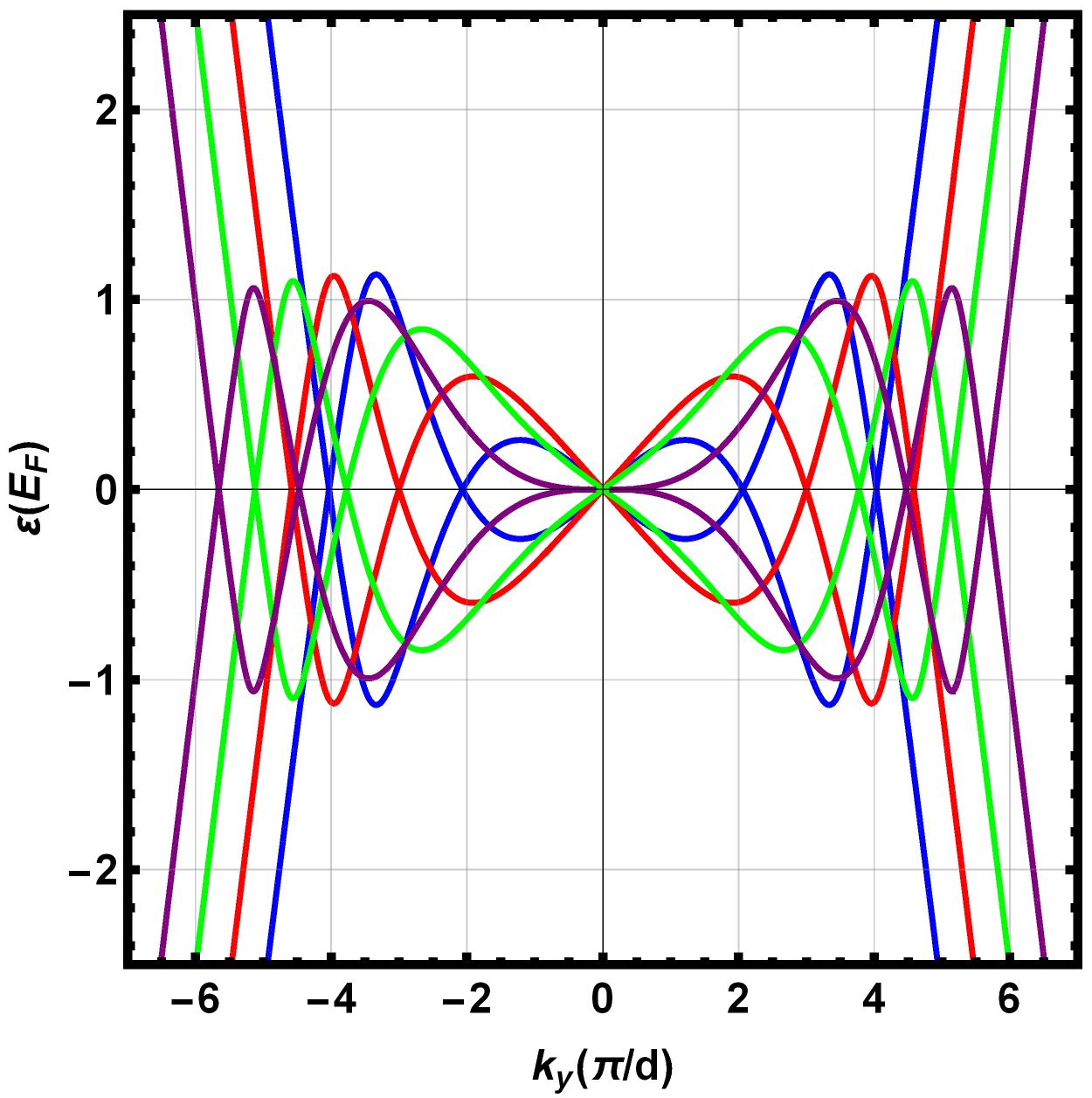}
    \label{FigCh33:SubFigC}
}
\caption{(Color online) Potential effect on the minibands of the electronic band structures. For $2\pi 
l\!<\!\mathbb{V}\!\leq\!2(l+1)\pi$ with $l\in \mathbb{N}$, $\mathbb{V}=2(l+\frac{1}{4})\pi$: blue color, 
$\mathbb{V}=2(l+\frac{1}{2})\pi$: red color, $\mathbb{V}=2(l+\frac{3}{4})\pi$: green color, and  $\mathbb{V}=2(l+1)\pi$: 
purple color. \protect\subref{FigCh33:SubFigD}, \protect\subref{FigCh33:SubFigE}, \protect\subref{FigCh33:SubFigF} for 
SSLGSL-3R with $q_2=\frac{1}{3}$. \protect\subref{FigCh33:SubFigA}, \protect\subref{FigCh33:SubFigB}, 
\protect\subref{FigCh33:SubFigC} for SSLGSL-2R. }
\label{FigCh33}
\end{figure}

After analyzing the potential effect on the minibands of the electronic band structures for  SSLGSL-3R with $q_2=\frac{1}{3}$ (Figures \ref{FigCh33:SubFigD}, \ref{FigCh33:SubFigE}, \ref{FigCh33:SubFigF}) and SSLGSL-2R (Figures \ref{FigCh33:SubFigA}, \ref{FigCh33:SubFigB}, \ref{FigCh33:SubFigC}), we summarize the following observations
\begin{itemize}
    \item SSLGSL-3R with $q_2=\frac{1}{3}$:
\begin{itemize}
      \item In Figure \ref{FigCh33:SubFigD} $(0\!<\!\mathbb{V}\!\leq\!2\pi)$, there is a single Dirac point located at the origin. All cones corresponding to different values of $\mathbb{V}$ are intersected in the original Dirac point (ODP).
      \item In Figure \ref{FigCh33:SubFigE} $(2\pi\!<\!\mathbb{V}\!\leq\!4\pi)$, there is only one Dirac point for all potentials where the opening gaps of its cone decrease with $\mathbb{V}$. For each potential, there is an emergence of two opening gaps, which decrease with  $\mathbb{V}$ moving away from  ODP.
      \item In Figure \ref{FigCh33:SubFigF} $(4\pi\!<\!\mathbb{V}\!\leq\!6\pi)$, there is always a single Dirac point centered at the origin as before, but we have appearance of four opening gaps, two for positive $k_{y}$ and two others for negative $k_{y}$.
      \item According to the above analysis, we conclude that in general way for any  potential such that $2\pi l\!<\!\mathbb{V}\!\leq\!2(l+1)\pi$ with $l\in \mathbb{N}$, one gets $2l$ opening gaps.
\end{itemize}
    \item SSLGSL-2R:
\begin{itemize}
      \item In Figure \ref{FigCh33:SubFigA} ($0\!<\!\mathbb{V}\!\leq\!2\pi$), there is a single Dirac point located at the origin where all cones are intersected. Gradually as $\mathbb{V}$ increases the cone opens and becomes parabolic at the value $\mathbb{V}=2 \pi$.
      \item In Figure \ref{FigCh33:SubFigB} ($2\pi\!<\!\mathbb{V}\!\leq\!4\pi$), for each value of $\mathbb{V}$ in addition to  ODP there is an emergence of tow additional Dirac points (ADPs), which are symmetric with respect to the original Dirac point.
      \item In Figure \ref{FigCh33:SubFigC} ($4\pi\!<\!\mathbb{V}\!\leq\!6\pi$), it is appeared four ADPs.
      \item We notice that if $2\pi l\!<\!\mathbb{V}\!\leq\!2(l+1)\pi$ with $l\in \mathbb{N}$, we have $2l$ ADPs. In this particular case, the opening gaps close to give ADPs located at $k_{y_{D_m}}\!=\!\sqrt{\mathbb{V}^2-(2m\pi)^2}/d$ (see Appendix).
\end{itemize}
\end{itemize}
\begin{figure}[!ht]
\centering
\subfloat[$l=0, \mathbb{V}=\pi$]{
    \includegraphics[scale=0.4]{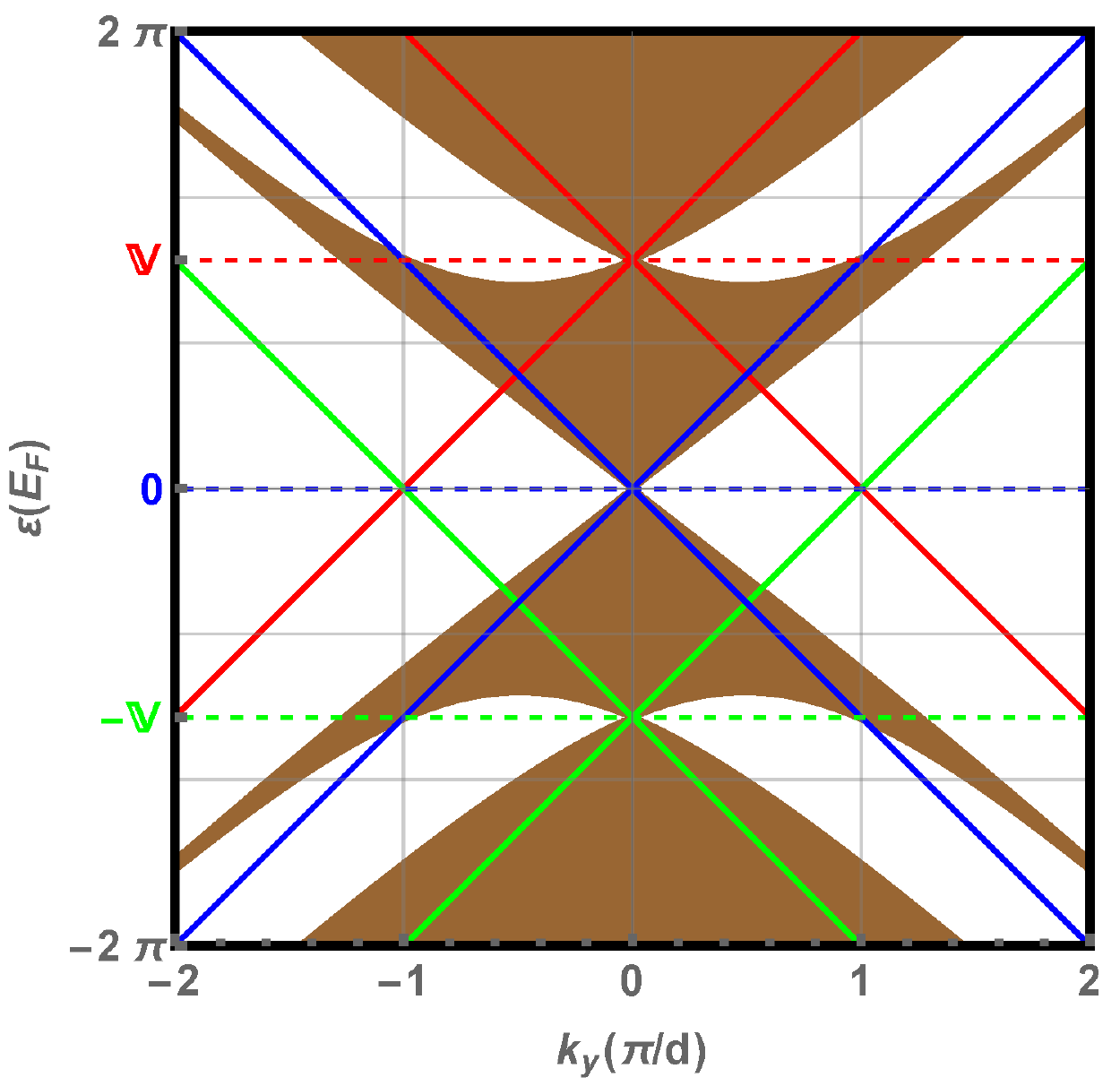}
    \label{FigCensus33:SubFigD}
}
\subfloat[$l=1, \mathbb{V}=3\pi$]{
    \includegraphics[scale=0.4]{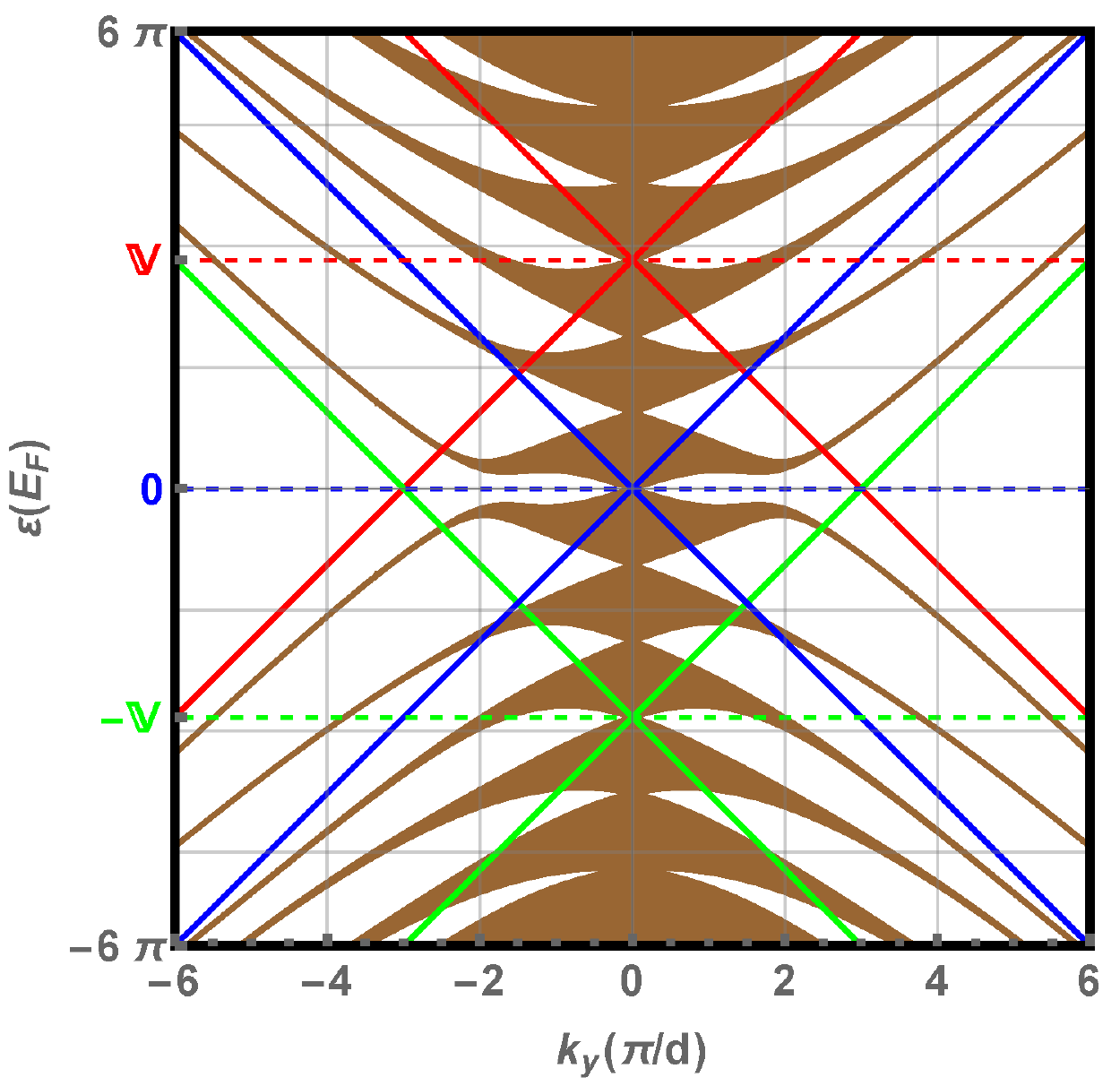}
    \label{FigCensus33:SubFigE}
}
\subfloat[$l=2, \mathbb{V}=5\pi$]{
    \includegraphics[scale=0.4]{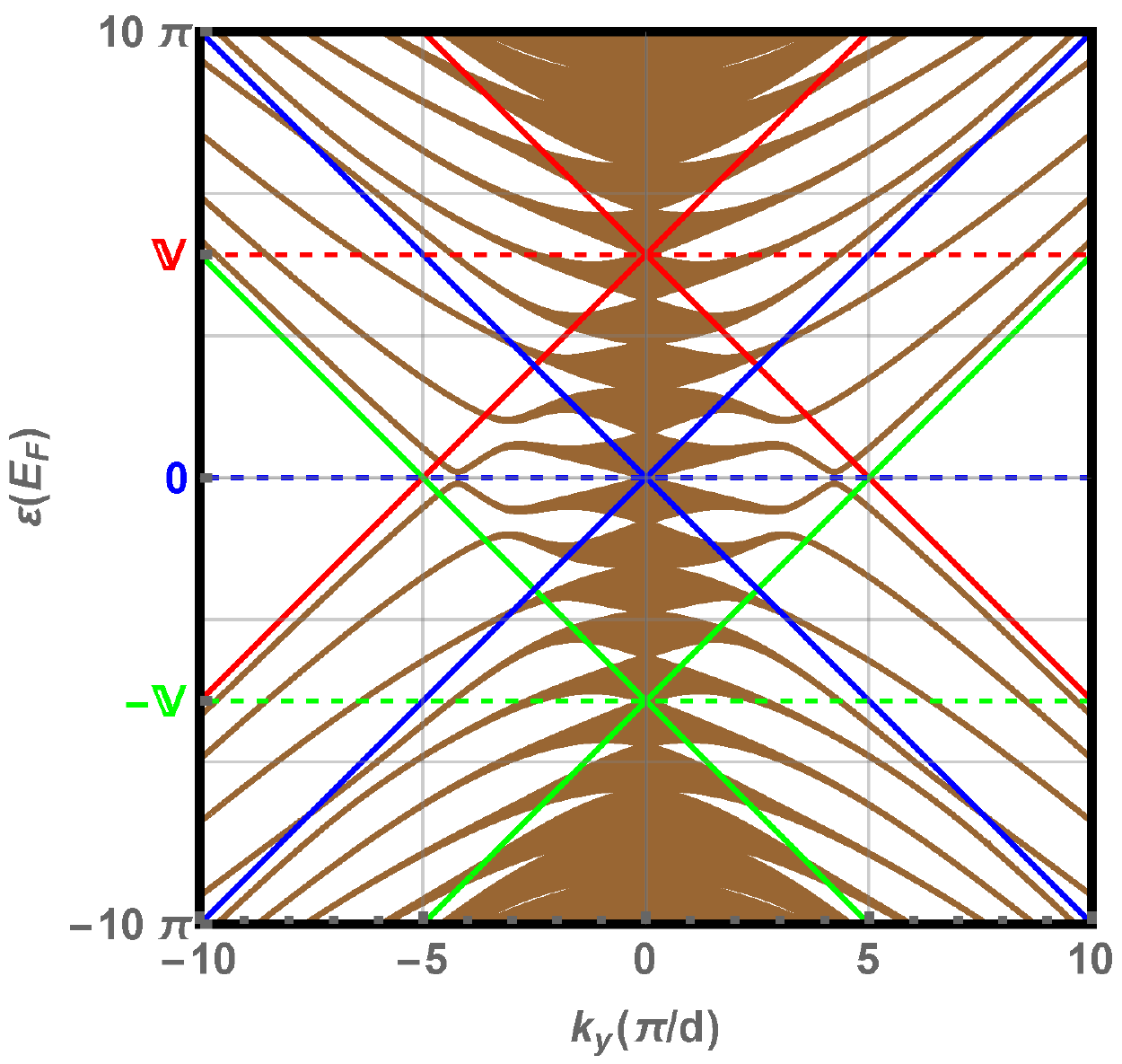}
    \label{FigCensus33:SubFigF}
}\\
\subfloat[$l=0, \mathbb{V}=\pi$]{
    \includegraphics[scale=0.4]{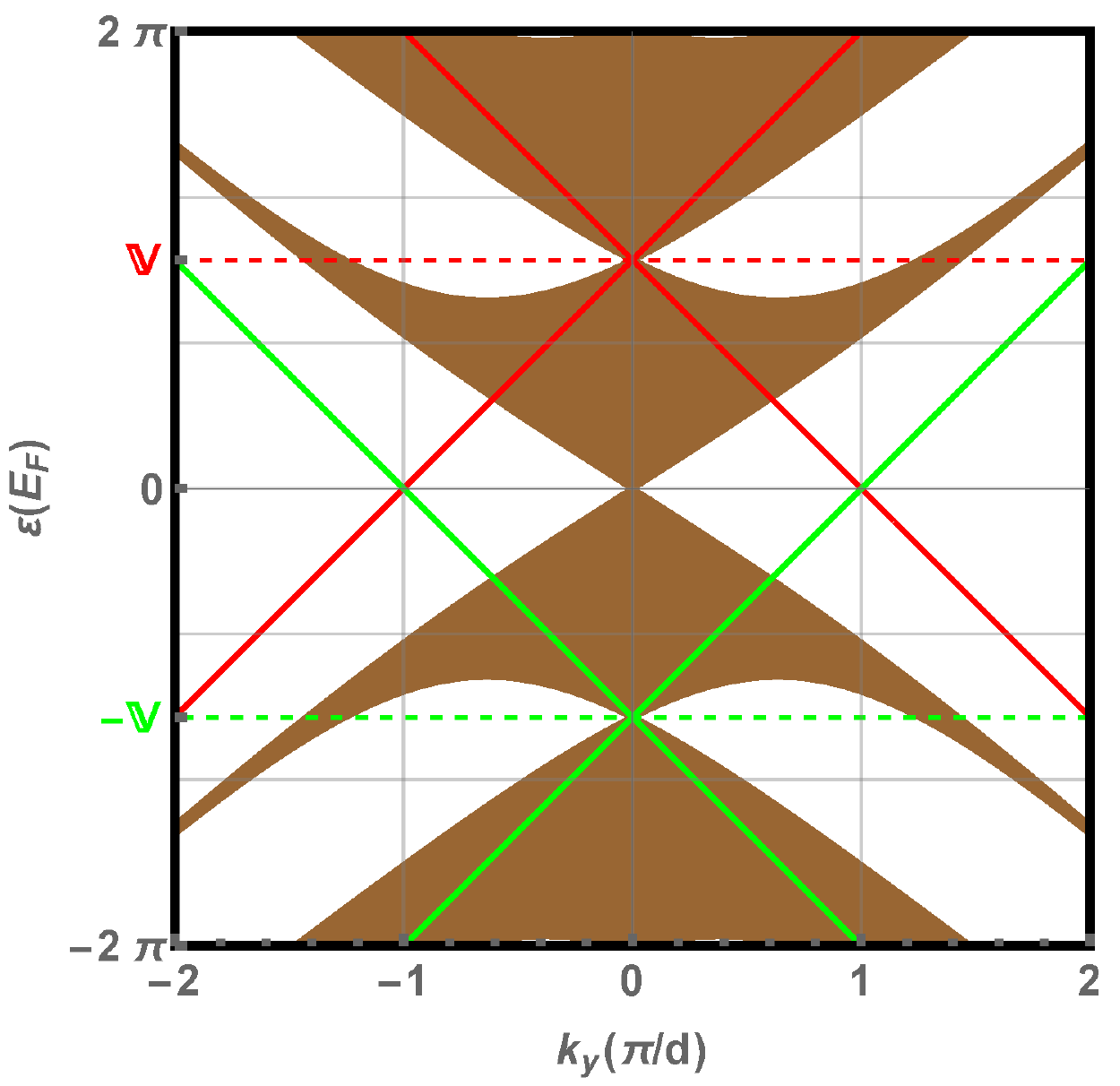}
    \label{FigCensus33:SubFigA}
}
\subfloat[$l=1, \mathbb{V}=3\pi$]{
    \includegraphics[scale=0.4]{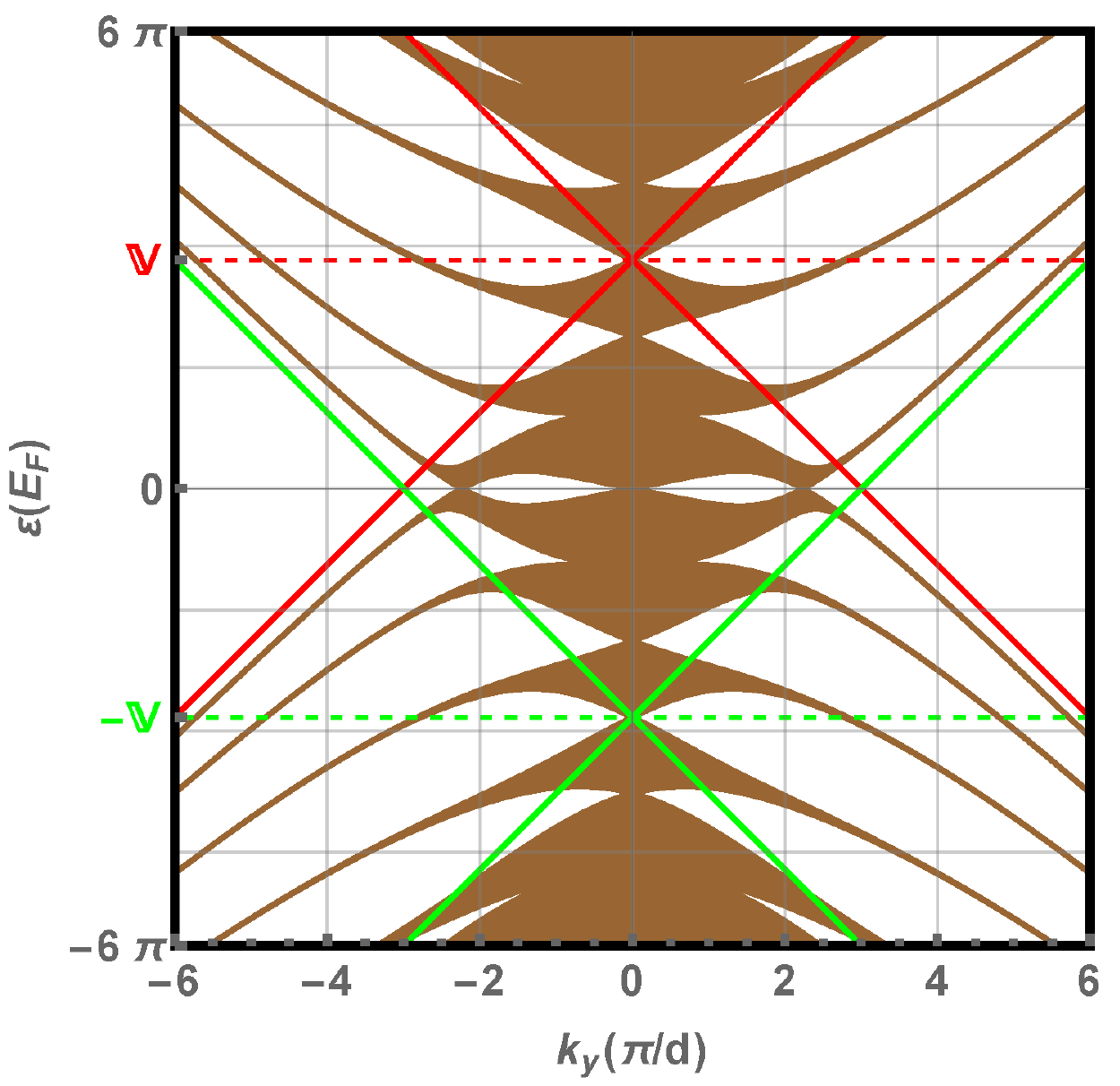}
    \label{FigCensus33:SubFigB}
}
\subfloat[$l=2, \mathbb{V}=5\pi$]{
    \includegraphics[scale=0.4]{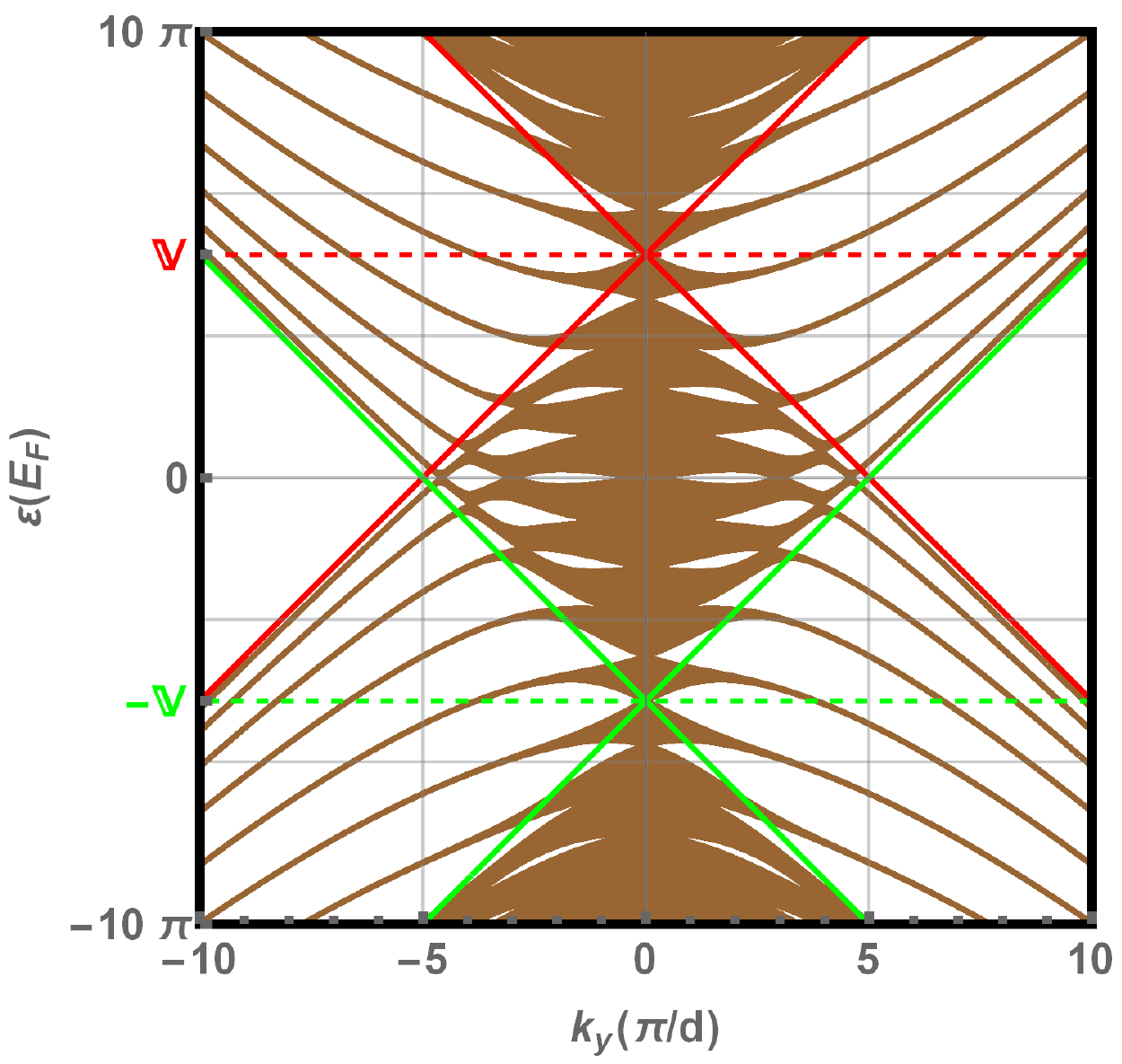}
    \label{FigCensus33:SubFigC}
}
\caption{(Color online) Inventory of vertical Dirac points in central zone and bound states in electronic band structures with $k_x\in[0,\frac{2\pi}{d}]$.
\protect\subref{FigCensus33:SubFigD}, \protect\subref{FigCensus33:SubFigE}, \protect\subref{FigCensus33:SubFigF} for SSLGSL-3R with $q_2=\frac{1}{3}$. \protect\subref{FigCensus33:SubFigA}, \protect\subref{FigCensus33:SubFigB}, \protect\subref{FigCensus33:SubFigC} for SSLGSL-2R.}
\label{FigCensus33}
\end{figure}

In the following, we enumerate the vertical Dirac points ($k_y=0$) (VDPs) and bound states in SSLGSL-3R and SSLGSL-2R where $2\pi l\!<\!\mathbb{V}\!\leq\!2(l+1)\pi$. For SSLGSL-3R, Figures \ref{FigCensus33:SubFigD}, \ref{FigCensus33:SubFigE}, \ref{FigCensus33:SubFigF} show that there are $(4l+3)$ VDPs (one of them is ODP) placed in the central zone $(-\mathbb{V}\!\leq\!\varepsilon\!\leq\!\mathbb{V})$, and each two consecutive one are separated by $\Delta\varepsilon\!=\!\pi$.
Between two consecutive VDPs, there are unbound states forming a lobe and therefore for $(4l+3)$ VDPs we have $2(2l+1)$ lobes. As long as $k_y$ is increased lobes are reduced to $(2l+1)$ bound states (non-parabolic and parabolic forms) in each quarter part of spectrum. For SSLGSL-2R, Figures \ref{FigCensus33:SubFigA}, \ref{FigCensus33:SubFigB}, \ref{FigCensus33:SubFigC} show that there is the same number of VDPs, lobes and bound states as for SSLGSL-3R. In this case, there is disappearance of non-parabolic bound states.

\begin{figure}[!ht]
\centering
  \includegraphics[scale=0.5]{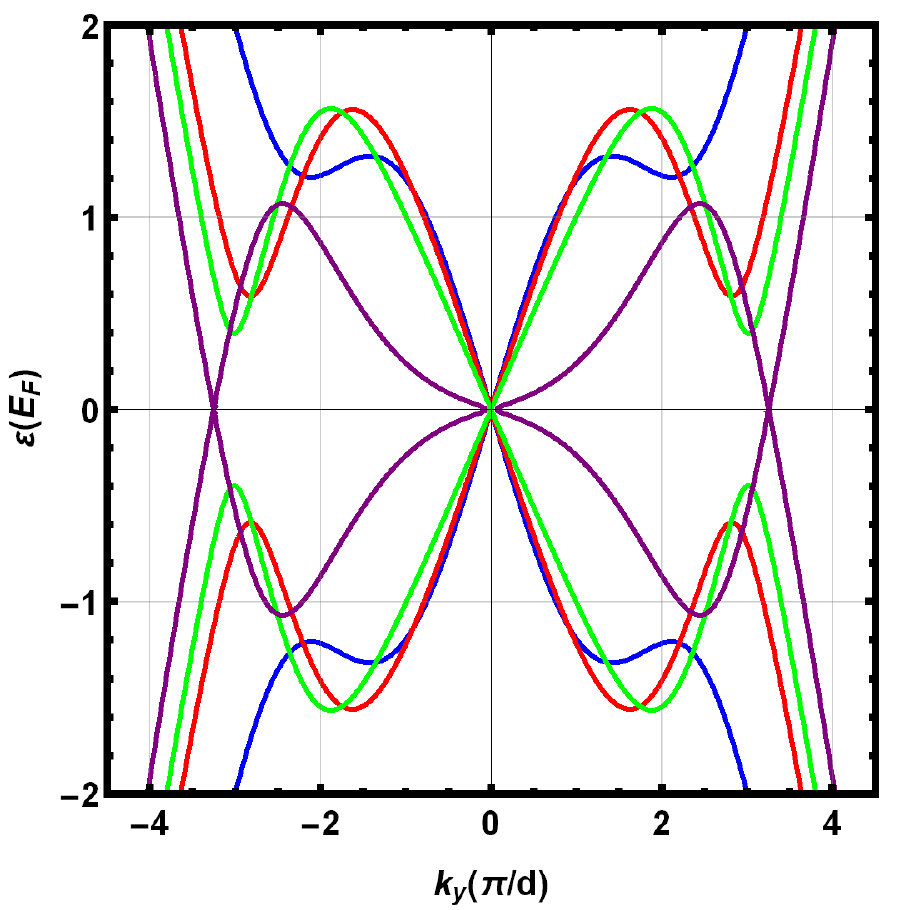}
\caption{(Color online) Effect of the distance $q_2$ on SSLGSL-3R for $k_x=0$, $\mathbb{V}=12$, $q_2=\frac{1}{2}$ (blue color), $q_2=\frac{1}{3}$ (red color),
$q_2=\frac{1}{4}$ (green color),
$q_2=0$ (purple color).}
\label{ch17}
\end{figure}
To highlight the role of the central region, we present in Figure \ref{ch17} the effect of the distance $q_2$ on SSLGSL-3R for $\mathbb{V}=12$, which is showing the opening gaps and its displacements towards  ODP. The opening gap decreases as long as $q_2$ decreases until a Dirac point is obtained. Two extreme cases occur: for  $q_{2}=\frac{d_2}{d}\longrightarrow 0$ our system behaves like SSLGSL-2R whereas for $q_{2}\longrightarrow 1$  it behaves like a pristine graphene with a single Dirac point at the origin.

\begin{figure}[!ht]
\centering
\subfloat[]{
    \includegraphics[scale=0.5]{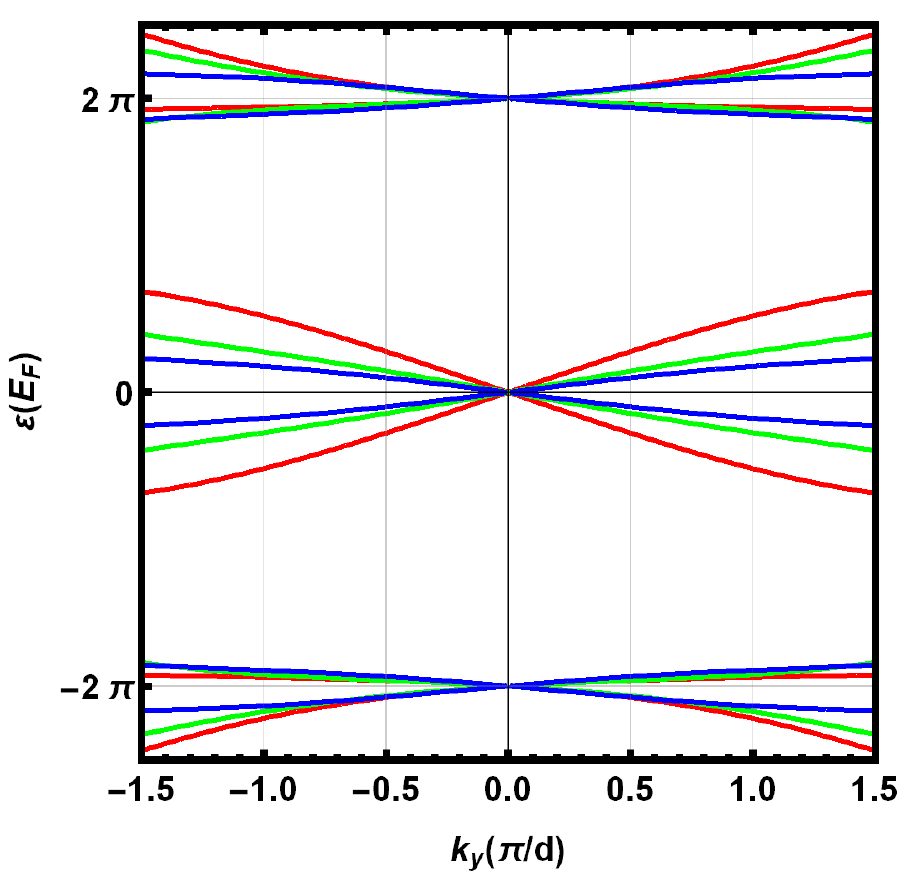}
    \label{FigDep:SubFigA}
}\hspace{10pt}
\subfloat[]{
    \includegraphics[scale=0.5]{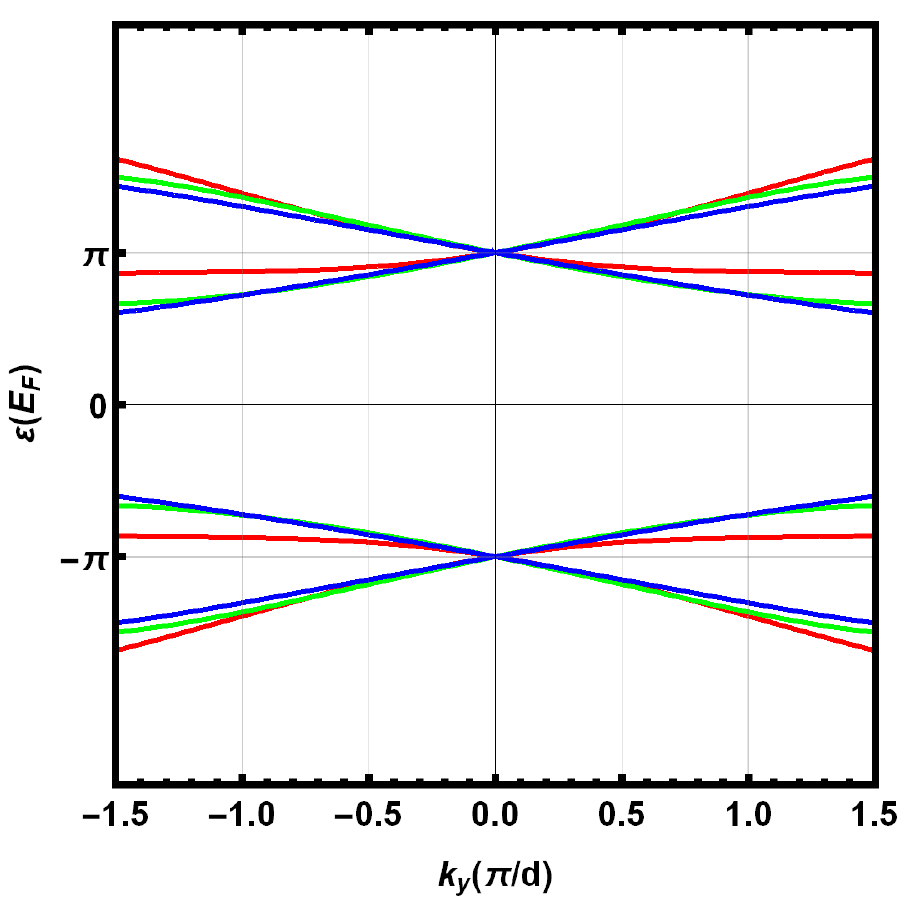}
    \label{FigDep:SubFigB}
}
\hspace{10pt}
\subfloat[]{
    \includegraphics[scale=0.555]{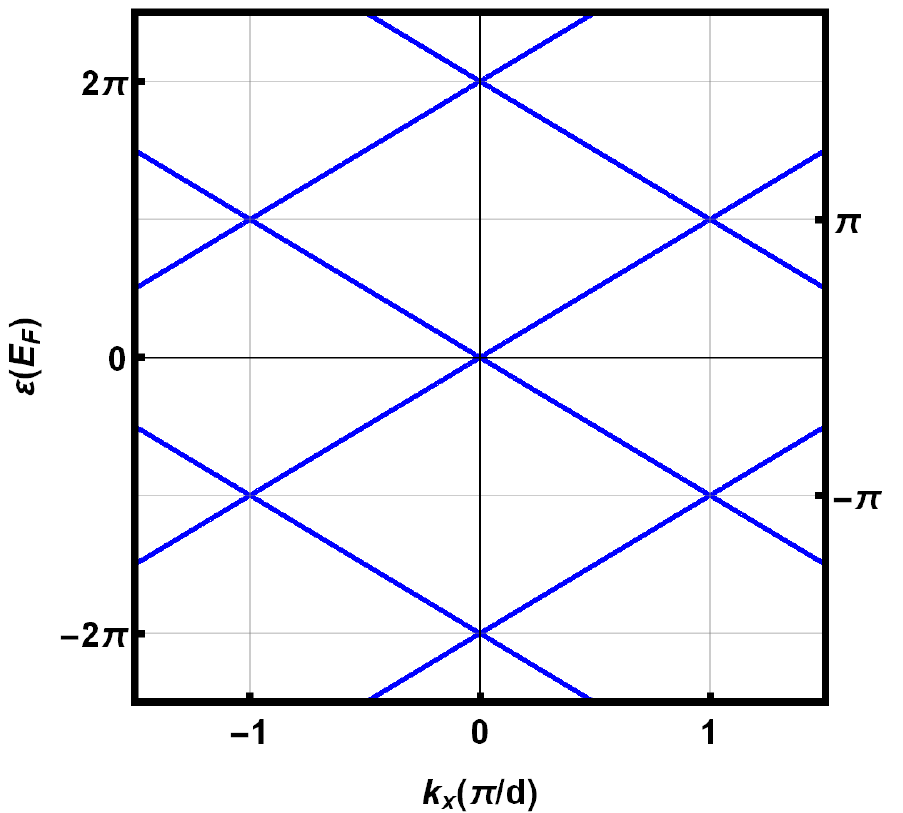}
    \label{FigDep:SubFigC}
}
\caption{(Color online) Location of VDPs in SSLGSL-3R for $q_2=\frac{1}{4}$ (blue curve), $q_2=\frac{1}{3}$ (green curve),  $q_2=\frac{1}{2}$ (red curve), $\mathbb{V}=5\pi$. \protect\subref{FigDep:SubFigA}: $k_x=0$. \protect\subref{FigDep:SubFigB}: $k_x=\pm\frac{\pi}{d}$.  \protect\subref{FigDep:SubFigC}: $k_y=0$ for any $\mathbb{V}$ and any $q_2$.}
\label{FigDep3}
\end{figure}
Figure \ref{FigDep:SubFigA} (\ref{FigDep:SubFigB}) presents, for $\mathbb{V}=5\pi$, the location of VDPs in the case $k_x = 0$ $(k_x=\pm\frac{\pi}{d})$ for SSLGLS-3R with three configurations of $q$. We observe the appearance of VDPs at energies $ 0, \pm2\pi$ $(\pm \pi)$, which are in agreement with the analytic results \eqref{EqDispl} (\eqref{EqDisplpid}) obtained in Appendix. The choice of potential amplitudes $(\mathbb{V}_1=-\mathbb{V}_3, \mathbb{V}_2=0)$ and the distances $(q_1=q_3)$ reduces (\ref{EqDispl}-\ref{EqDisplpid}) to the following energies
\begin{equation}\label{430}
\varepsilon\left(k_x=0\right)= 2m\pi,\qquad
\varepsilon\left(k_x=\frac{\pi}{d}\right)=\varepsilon\left(k_x=-\frac{\pi}{d}\right)=(2m\pm 1)\pi.
\end{equation}
Now, it is clear that the appearance of VDPs is corresponding to take the configuration $ m=0,\pm 1 $ $ (m=0) $. Figure \ref{FigDep:SubFigC} presents the VDPs location for $k_y=0$, and we observe that the band structures are independent on the parameters $\mathbb{V}$ and $q_2$.
\begin{figure}[!ht]
\centering
\subfloat[]{
   \includegraphics[scale=0.55]{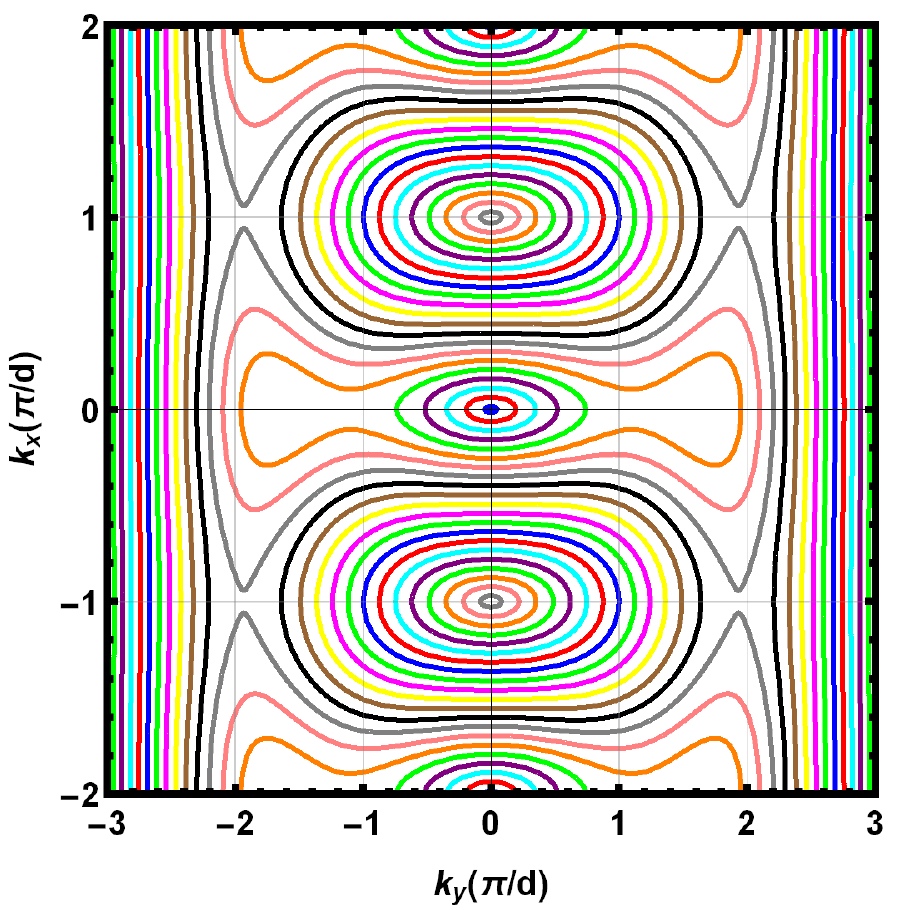}
    \label{FigAniso1:SubFigB}
}\hspace{20pt}
\subfloat[]{
    \includegraphics[scale=0.55]{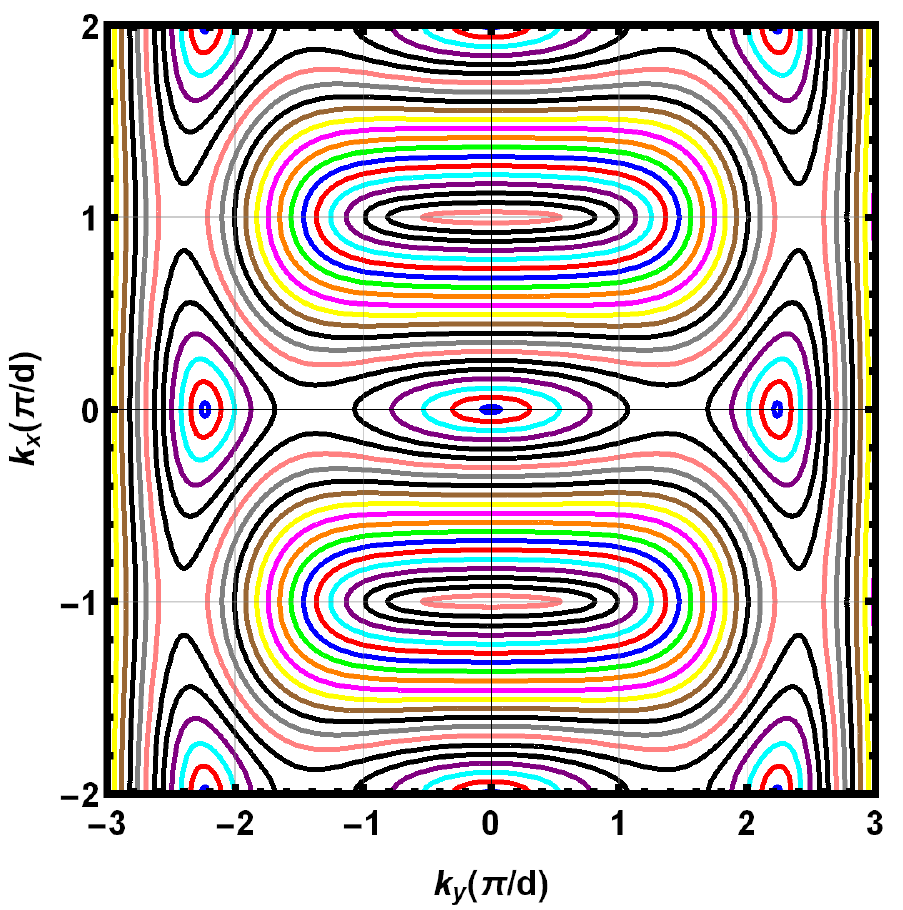}
    \label{FigAniso1:SubFigA}
}
\caption{(Color online) Contour plot of the energy band $\varepsilon\in[0,3]$ with step $0.15$, $\mathbb{V}=3\pi$.
\protect\subref{FigAniso1:SubFigB}: SSLGSL-3R with $q_2=\frac{1}{3}$, \protect\subref{FigAniso1:SubFigA}: SSLGSL-2R.}
\label{FigAniso1}
\end{figure}

Figure \ref{FigAniso1} shows the contour plot of the low energy band around the Dirac points for two configurations of SSLGSL-3R and SSLGSL-2R with $\mathbb{V}=3\pi$. The choice of such value of $\mathbb{V}$ allows us to have an ODP for SSLGSL-3R (Figure \ref{FigAniso1:SubFigB}) and three ADPs for SSLGSL-2R (Figure \ref{FigAniso1:SubFigA}). Each panel of the Figure \ref{FigAniso1} contains two Brillouin zones where there is closed (not closed) energies contours corresponding to the unbound (bound) states. We notice that there is an anisotropy of contours  near Dirac points, in fact we have a clear difference between the contours at  ODP (ellipse-like) and those near the additional ones (fewer symmetric contours on two sides). At some level of energy, the contours of the three Dirac points merge by giving a single contour in the form of a dumbbell, which envelops the three Dirac points. To understand the energy dispersion near Dirac points, we write \eqref{Diracdispersion} around  ODP for SSLGSL-3R as
\begin{equation}\label{Diracdispertion}
  \varepsilon=\pm d \sqrt{k_x^2+\frac{1}{\mathbb{V}^2}\left(2+q_2^2\mathbb{V}^2-2\cos((q_2-1)\mathbb{V})
  -2q_2\mathbb{V}\sin\left((q_2-1)\mathbb{V}\right)\right)k_y^2}
\end{equation}
which describes the elliptical contours of the energies in the vicinity of  ODP for SSLGSL-3R with $q_2=\frac{1}{3}$ in Figure \ref{FigAniso1:SubFigB} and SSLGSL-2R in Figure \ref{FigAniso1:SubFigA}. In order to explain the difference between the contour at the ODP and those at  ADPs in Figure \ref{FigAniso1:SubFigA} for SSLGSL-2R, we have to find the dispersion relation around  ADPs. Indeed, writing \eqref{Diracdispersion} in the vicinity of $(k_x=0,k_y=k_{y_{D_m}},\varepsilon=0)$, we obtain
\begin{equation}\label{Eq50}
  \varepsilon_m=\pm d\sqrt{k_x^2+\left[1-\left(\frac{2m\pi}
  {\mathbb{V}}\right)^2\right]^2\left(k_y-k_{y_{D_m}}\right)^2}.
\end{equation}
For Figure \ref{FigAniso1:SubFigA}, \eqref{Eq50} can be written as
\begin{equation}\label{Eq51}
\varepsilon_1=\pm d\sqrt{k_x^2+\left(\frac{5}{9}\right)^2\left(k_y-\frac{\sqrt{5}\pi}{d}\right)^2}.
\end{equation}
\eqref{Diracdispersionyyy} in Appendix gives the dispersion relation in the vicinity of VDP $(k_x=\frac{\pi}{d}, k_y=0, \varepsilon= \pi)$
\begin{equation}
  \varepsilon-\pi=\pm \sqrt{d^2(k_x-\frac{\pi}{d})^2+B_2(\pi)k_y^2}
\end{equation}
as well as that in the vicinity of VDP $(k_x=-\frac{\pi}{d}, k_y=0, \varepsilon= \pi)$
\begin{equation}
  \varepsilon-\pi=\pm \sqrt{d^2(k_x+\frac{\pi}{d})^2+B_2(\pi)k_y^2}
\end{equation}
which explains the energy contours around the points $(k_x=\pm\frac{\pi}{d}, k_y=0, \varepsilon= \pi)$ in Figure \ref{FigAniso1}.
\begin{figure}[!ht]
\centering
\subfloat[]{
    \includegraphics[scale=0.55]{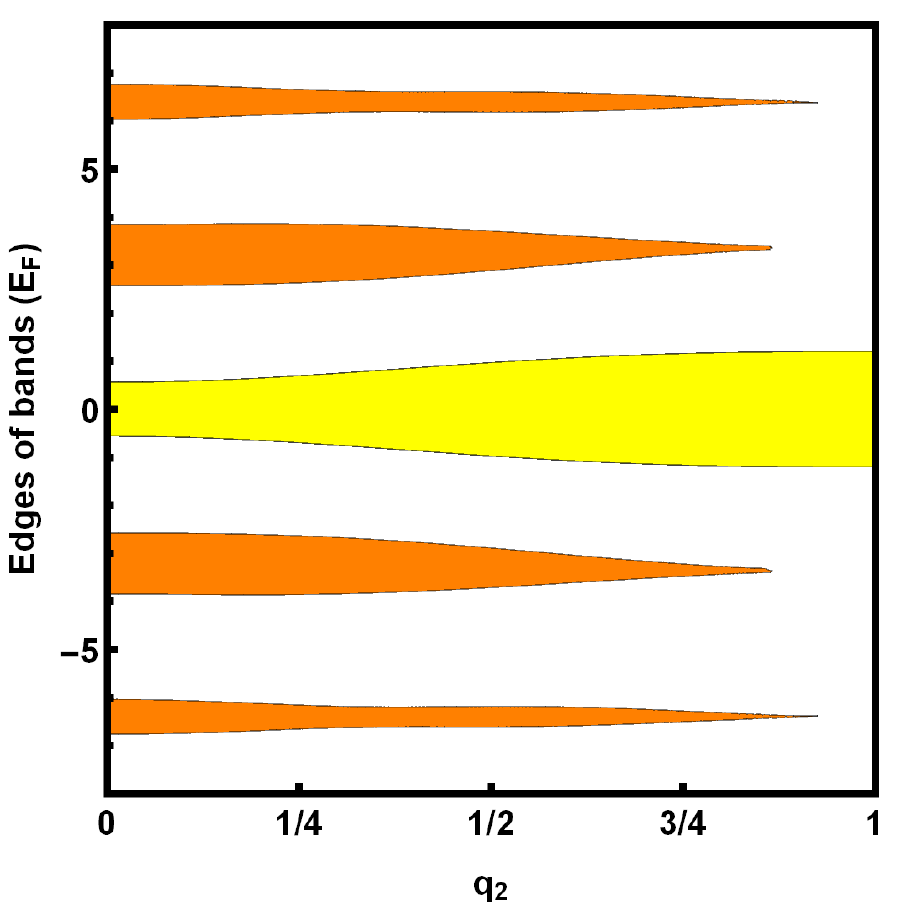}
    \label{FigEdges:SubFigC}
}\hspace{20pt}
\subfloat[]{
    \includegraphics[scale=0.55]{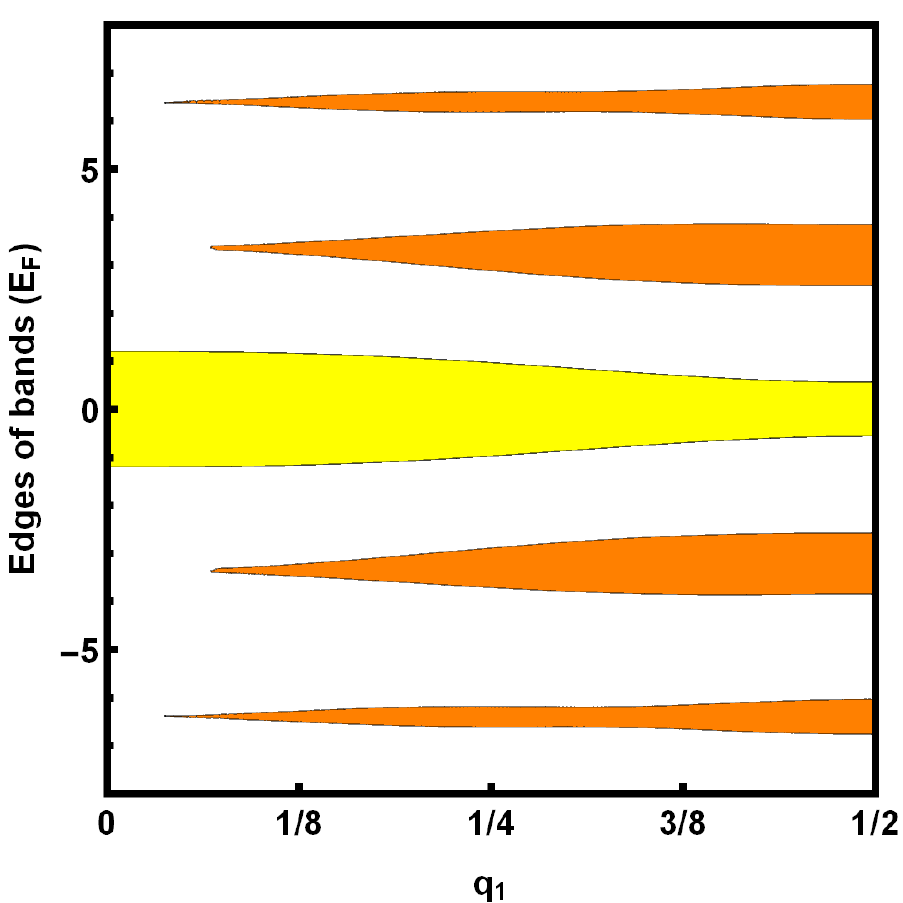}
    \label{FigEdges:SubFigA}
}
\caption{(Color online) Evolution of mini-band edges for $\mathbb{V}=4$ and $k_y=0.12~\nano\meter^{-1}$. \protect\subref{FigEdges:SubFigC}: $q=\left\{\frac{1 - q_2}{2}, q_2, \frac{1 - q_2}{2}\right\}$. \protect\subref{FigEdges:SubFigA}: $q=\{ q_1, 1-2q_1, q_1\}$ with $0\leq q_1\leq\frac{1}{2}$.}
\label{Edges}
\end{figure}

In Figure \ref{FigEdges:SubFigC} (\ref{FigEdges:SubFigA}), we illustrate the evolution cutting slices between the energy cones and the plane $k_y=0.12~\nano\meter^{-1}$ in similar way as in \cite{Pham2010}, for SSLGSL-3R  with configuration $q=\{\frac{1 - q_2}{2}, q_2, \frac{1 - q_2}{2}\}$ $(q=\{q_1, 1-2 q_1, q_1\}, 0\!\leq\! q_1\!\leq\!\frac{1}{2})$. We choose $\mathbb{V}\!=\!4$ so that there is no opening gaps and $k_y=0.12~\nano\meter^{-1}$ to stay widely near to ODP. When $ q_2$ $( q_1)$ varies from $0$ to $1$ $\left(\frac{1 }{2}\right)$, the system gradually changes from a SSLGSL-2R (pristine graphene) to a pristine graphene (SSLGSL-2R) through  SSLGSLs-3R. The yellow slice at  mini-bands, called "cutoff region", has the width of $|2 d k_y|$ on the side of $q_2=1$ $(q_1=0)$, i.e. when the system becomes pristine graphene. The width of the yellow slice is smaller on the side of SSLGSL-2R. Away from mini-bands, the orange slices are the band gaps near VDPs in the bound states zone (brown zone in Figure \ref{StructAband1}). When $q_2$ $(q_1)$ increases (decreases), the width of band gaps decrease until their disappearance when the system becomes pristine graphene. When $q_2\longrightarrow 1$  $(q_2\longrightarrow 0)$, the band structures contain only  ODP with a purely linear dispersion. Figures \ref{FigEdges:SubFigC}, \ref{FigEdges:SubFigA} show a mirror symmetry between them. The "cutoff region" and the band gaps in SSLGSL-3R are horizontal unlike  those corresponding to SSLGSL-2R \cite{Pham2010}.

%======================================================================
\subsection{Normalized group velocity}
%======================================================================

\begin{figure}[!ht]
\centering
\subfloat[]{
    \includegraphics[scale=0.7]{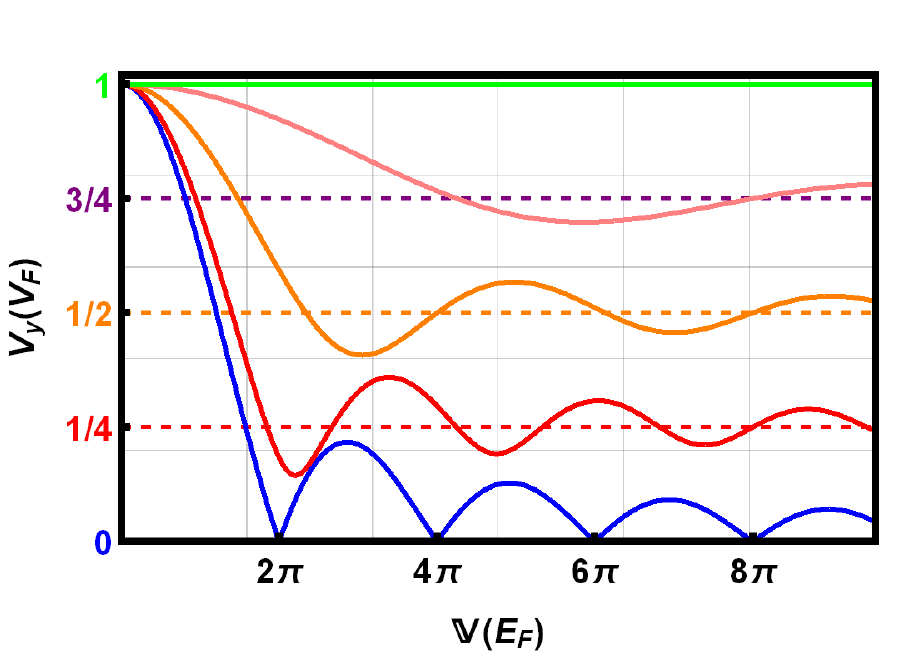}
    \label{FigVG0:SubFigB}
}\hspace{20pt}
\subfloat[]{
    \includegraphics[scale=0.7]{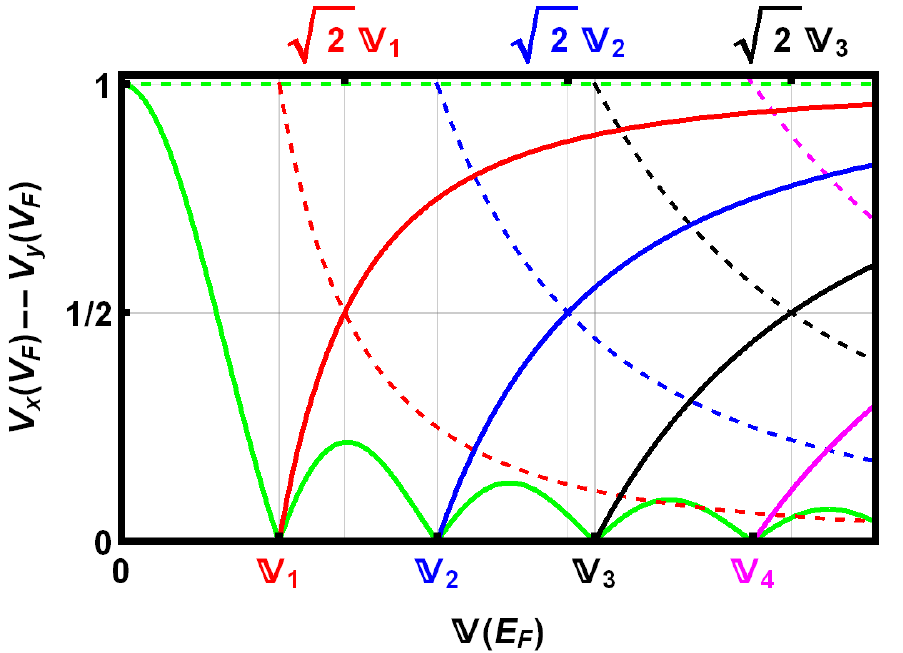}
    \label{FigVG0:SubFigA}
}
\caption{(Color online) \protect\subref{FigVG0:SubFigB}: For SSLGSL-3R, normalized velocity $\frac{v_y}{v_F}$ versus potential amplitude $\mathbb{V}$ for states with $q_2=0$ (solid blue line), $q_2=\frac{1}{4}$ (solid red line), $q_2=\frac{1}{2}$ (solid orange line), $q_2=\frac{3}{4}$ (solid pink line) and $q_2=1$ (solid green line). \protect\subref{FigVG0:SubFigA}: For SSLGSL-2R, normalized velocities $\frac{v_x}{v_F}$ and $\frac{v_y}{v_F}$ versus potential amplitude $\mathbb{V}$ for: state near  ODP (dashed and solid green line), state near the first ADP (dashed and solid red line), state near the second ADP (dashed and solid blue line), state near the third ADP (dashed and solid black line), and state near the fourth ADP (dashed and solid magenta line). ADPs appear at $k_{y_{D_m}}\!=\!\sqrt{\mathbb{V}^2-\mathbb{V}_m^2}/d$ where $\mathbb{V}_m=2 m\pi$ with $m\in \mathbb{N}^{*}$.}
\label{VG0}
\end{figure}

For SSLGSL-3R, we can derive the two components of the normalized group velocity near  ODP from \eqref{GV3} (see Appendix). Thus, they are given by
\begin{equation}\label{GVS3}
  \frac{v_x}{v_F}=1, \qquad
  \frac{v_y}{v_F}=\frac{1}{\mathbb{V}}
  \sqrt{2+q_2^2\mathbb{V}^2-2\cos\left(\left(q_2-1\right)\mathbb{V}\right)
  -2q_2\mathbb{V}\sin\left(\left(q_2-1\right)\mathbb{V}\right)}.
\end{equation}
In Figure \ref{FigVG0:SubFigB}, we represent the normalized group velocity $\frac{v_y}{v_F} $ versus the potential amplitude $\mathbb{V}$ with some values of $q_2$ for the electronic band structures in the vicinity of ODP. According to Figure \ref{FigVG0:SubFigB}, we summarize the following interesting results:
\begin{itemize}
	\item When $q_2$ increases, the amplitude of oscillations  of $\frac{v_y}{v_F} $ decreases and its periodicity increases as long as $\mathbb{V}$ increases. Up to a big value of $\mathbb{V}$ we end up with $\lim\limits_{\mathbb{V} \longrightarrow +\infty} \frac{v_y}{v_F} =q_2$.
	\item  For $q_2 \neq 0$, the $\frac{v_y}{v_F} $ values oscillate alternately positively and negatively  around the distance $q_2$.
	\item For $q_2=0$, $\frac{v_y}{v_F} $ presents an inversion of the negative alternations, the curve is redressed and it oscillates only in positive alternation.
\end{itemize} 

Recall that SSLGSL-2R is a particular case of SSLGSL-3R, and the corresponding two components of the normalized group velocity near  ODP can be obtained from \eqref{GVS3} by fixing $q_2 = 0$
\begin{equation}\label{GVS20}
  \frac{v_x}{v_F}=1,\qquad \frac{v_y}{v_F}=\frac{\sqrt{2}}{\mathbb{V}} \sqrt{1-\cos (\mathbb{V})}
\end{equation}
SSLGSL-2R has ADPs located at ($k_x=0, k_y=k_{y_{D_m}},\varepsilon=0$) and the normalized group velocities near such ADPs are given by
\begin{equation}\label{GVS20add}
  \frac{v_x}{v_F}=\left(\frac{2m\pi}{\mathbb{V}}\right)^2, \qquad
  \frac{v_y}{v_F}=1-\left(\frac{2m\pi}{\mathbb{V}}\right)^2.
\end{equation}
For SSLGSL-2R, in Figure \ref{FigVG0:SubFigA} we plot the normalized group velocities $\frac{v_x}{v_F} $ (dashed lines) and $\frac{v_y}{v_F}$ (solid lines) versus the potential amplitude $\mathbb{V}$. In the vicinity of ODP, the variation of $\frac{v_y}{v_F}$ is represented by the solid green line, which has a half crest between $ 0 $ and $ 2 \pi $. It decreases from $1$ to $0$ and then a succession of crests, which appear between each $2\pi$ whose amplitude decreases as long as $\mathbb{V}$ increases. The green dashed line shows that $v_x=v_F$. First, we notice that the values where $\frac{v_y}{v_F} $ of  ODP is null coincide with those where  ADPs appear, namely $\mathbb{V}_m=2 m\pi$ with $m\in \mathbb{N}^{*}$. This coincidence proves the existence of a relation between the appearance of ADPs and the anisotropy of the group velocity. In addition, the new ADPs contribute to the flow of charge carriers once they appear. We plot $\frac{v_x }{v_F}$ and $\frac{v_y}{v_F} $ for states with $k_y=k_{y_{D_m}}$ near the $m$-th ADP for some first values of $m$ by the red, blue, black and magenta lines for $m=1,2,3$ and $4$, respectively. To understand the transport characteristics of  SSLGSL-2R, it is very important to notice that there is no potential $\mathbb{V}$, which can cancel all these velocities at the same time, except for the potential $\mathbb{V}=\mathbb{V}_1$ associated with the emergence of the first pair of new Dirac points. We observe that at $\mathbb{V}_m$, where one pair of Dirac points are emerging, $\frac{v_x}{v_F} $  decreases from $1$ but $\frac{v_y}{v_F} $  increases from zero giving the sum $\frac{v_x}{v_F} + \frac{v_y}{v_F}=1$. When the two components $\frac{v_x}{v_F} = \frac{v_y}{v_F} = \frac{1}{2}$  and now from \eqref{GVS20add}, we get the relation $\mathbb{V}(v_x = v_y) =\sqrt{2}\mathbb{V}_m $, which has been illustrated in the Figure \ref{FigVG0:SubFigA} generated from the dispersion relation \eqref{33}
and also has been already obtained in \cite{Barbier2010,Pham2010}.

In the vicinity of Dirac points $(k_x=\pm\frac{\pi}{d}, k_y=0, \varepsilon= \pi)$ and $(k_x=\pm\frac{\pi}{d}, k_y=0, \varepsilon= -\pi)$, \eqref{GV3} in Appendix gives the normalized group velocity
\begin{equation}\label{GVedges}
  \frac{v_x}{v_F}=1,\qquad \frac{v_y}{v_F}=\frac{1}{d} \sqrt{B_2(\pi)}
\end{equation}
with the condition $\mathbb{V}\neq \pi$.
\begin{figure}[!ht]
\centering
   \includegraphics[scale=0.6]{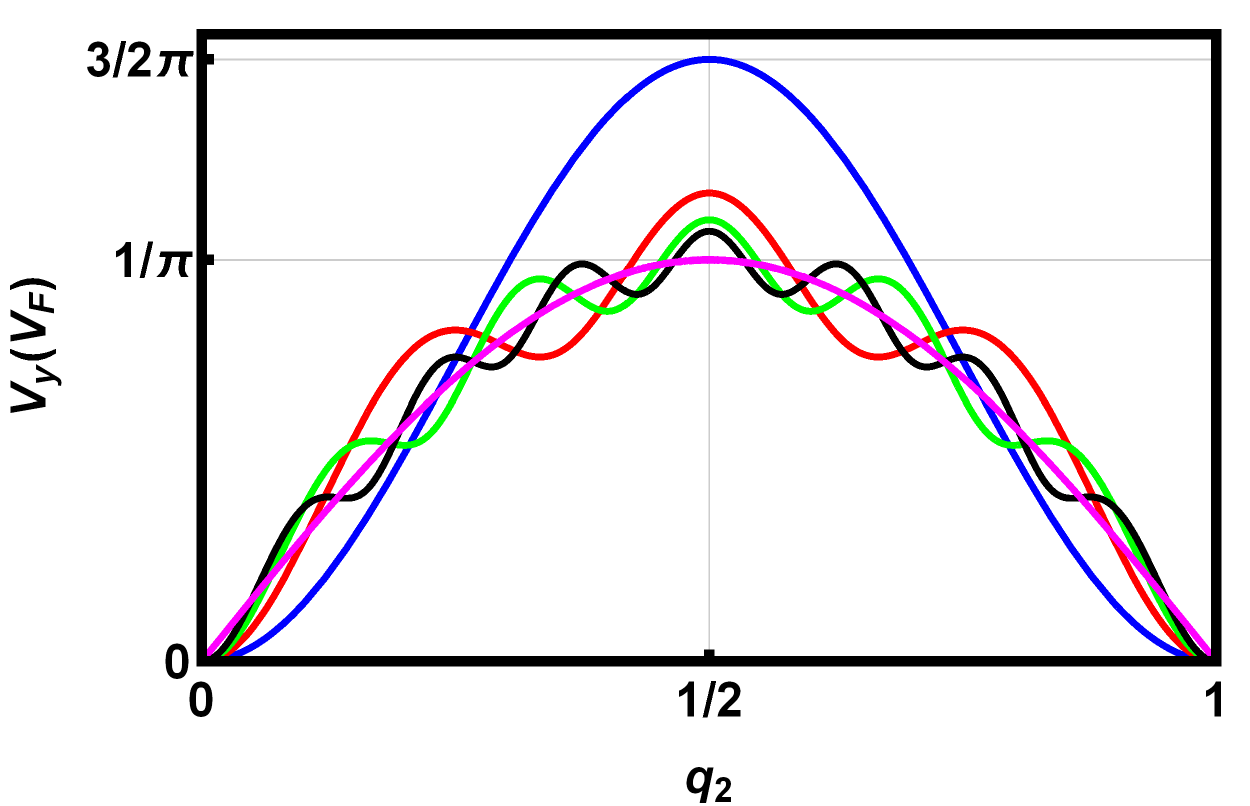}
\caption{(Color online) Group velocity $\frac{v_y}{v_F}$ near Dirac point $(k_x=\frac{\pi}{d}, k_y=0, \varepsilon= \pi)$ for $\mathbb{V}=(4p+3)\pi$ versus $q_2$ with $p=0, 1, 2, 3, \infty$ corresponding to blue, red, green, black and magenta curves, respectively.}
\label{Moh}
\end{figure}

Figure \ref{Moh} elucidates the normalized group velocity $\frac{v_y}{v_F}$ versus $q_2$ for $\mathbb{V}=(4p+3)\pi$ with $p$ is an integer. It shows that $\frac{v_y}{v_F}$ is maximum at $q_2=\frac{1}{2}$ and we can write
\begin{equation}
  \left.\frac{v_y}{v_F}\right)_{q_2=\frac{1}{2}}=\frac{(4 p+3) \sqrt{4 p (2 p+3)+(4 p+3) \cos (2 \pi  p)+5}}{4 \sqrt{2} \pi  \left| (p+1) (2 p+1)\right| }.
\end{equation}
All maximum values of $\frac{v_y}{v_F}$ are between two limit values $\frac{1}{\pi}$ and $\frac{3}{2\pi}$ corresponding to $p\longrightarrow 0$ and $p\longrightarrow \infty$, respectively.

%======================================================================
\section{Conclusion}\label{Sec:Conclusion}
%======================================================================

In the continuum model based on an effective Dirac equation, by using the transfer matrix method and Bloch theorem, we have determined the dispersion relation for single layer graphene superlattice with three regions (SLGSL-3R). These systems can be realized, for example, by applying an appropriate external periodic potentials. The elementary cell of our SSLGSL-3R, composed of three successive regions of potential $\mathbb{V}$, $0$, and $-\mathbb{V}$, and characterized by $\mathbb{V}$, $d$, and $q=\left\{\frac{1 - q_2}{2}, q_2, \frac{1 - q_2}{2}\right\}$.

A detailed theoretical study, of different Dirac point locations (ADPs, VDPs) and normalized  group velocity near them, was the object of our appendix. Indeed, we were particularly interested to the horizontal and vertical contact points (HCPs, VCPs) in the energy spectrum corresponding to $\varepsilon=0$ and $k_y=0$, respectively. The vertical contact points (VCPs) in the first Brillouin zone and horizontal contact points (HCPs) in minibands corresponding to the cases $(k_x=\pm\frac{\pi}{d},k_y=0)$ and $(k_x=0,\varepsilon=0)$, respectively. There were implemented in the dispersion relation \eqref{33} to obtain the energies (\ref{EqDispl}-\ref{EqDisplpid}) related to VCPs  location and \eqref{ADP} relating to the HCPs  location. Before applying Taylor's approximation formula, we have transformed our dispersion relation into an implicit function  $f(k_x ,k_y,\varepsilon )=0$ where its gradient is zero at (HCPs, VCPs) and its Hessian matrices are diagonal. The application of the Taylor formula, near the contact points, gave a quadratic relation \eqref{quadratic}, which has been used to derive an energy with Dirac-like form and allowed us to conclude that the considered (HCPs, VCPs) are additional Dirac points (ADPs) and vertical Dirac points (VDPs), respectively. The components of the normalized group velocity, in the vicinity of the Dirac Points were also derived from \eqref{quadratic}.

Interesting numerical results concerning the SSLGSL-3R electronic band structures have been reported. It was shown that the electronic band structures contain zones depending on the number of cones in the elementary cell. In the general case of SSLGSL-3R, we have obtained five different zones: forbidden and of unbound, non-parabolic bound, serried parabolic bound, and less serried parabolic bound states. In the particular case of SSLGSL-2R, there is only parabolic states, the five zones are reduced to the three zones: forbidden, bound and unbound states. The potential amplitude $\mathbb{V}$ has a very remarkable effect in minibands, because we have found the appearance of opening gaps (ADPs) in SSLGSL-3R (SSLGSL-2R) as long as $\mathbb{V}$ increased. We have noticed that ODP exists for all SSLGSL-3R configurations. The number of opening gaps and ADP depend also on the applied potential $\mathbb{V}$, indeed for $\mathbb{V}\in]2l\pi,2(l + 1)\pi]$ there are in the central zone ($4l+3$) VDPs (one of them is ODP), $2l$ opening gaps (ADPs) in SSLGSL-3R (SSLGSL-2R). Gradually as long as the distance $q_2$ decreases, the opening gaps in minibands shrink and move away from  ODP until appearance of ADPs at $k_{y_{D_m}}=\pm\frac{1}{d}\sqrt{\mathbb{V}^2-(2m\pi)^2}$ with ($\mathbb{V}^2>(2m\pi)^2$ and $ m \in \mathbb{N}^{*}$)  where $q_2=0$. VDPs are located at $ \varepsilon= m\pi$ with $m \in \mathbb{Z}$,  which takes place when $\cos(k_x d)=\pm 1$. 

The contour plot of the energy band was illustrated near Dirac points, and we have 
observed that the closed (open) contours depict unbound (bound) states and just 
near  DPs the contours have a like-elliptical form.  Subsequently, 
we have illustrated the evolution cutting slices between the energy cones and a plane just near ODP, the transition between the limiting cases of $q_2\longrightarrow 0$ $\left(q_1\longrightarrow \frac{1}{2}\right)$ and $q_2\longrightarrow 1$ ($q_1\longrightarrow 0$)  the system  changes from a SSLGSL-2R to a pristine graphene through SSLGSLs-3R. It was shown that the width of ”cutoff region” is $|2 d k_y|$ and there is only ODP on the side of $q_2=1$ $(q_1=0)$.

A numerical exploitation of the normalized group velocity corresponding to SSLSL-3R
has been also reported. It was shown that the normalized velocity near ODP of a SSLSL-3R for different $q_2$$,  \frac{v_x}{v_F}=1$ and $\lim\limits_{\mathbb{V}\longrightarrow +\infty} \frac{v_y}{v_F} =q_2$. Our general  SSLSL-3R  dispersion relation, allowed us to reproduce the behavior of the SSLSL-2R group velocity near (ODP, ADPs) that was found in \cite{Pham2010,Barbier2010}. Finally, we have studied an interesting case of  the group velocity
 near VDP $\left(k_x=\pm\frac{\pi}{d}\right)$ versus $q_2$ for $\left(\mathbb{V}=(4p+3)\pi,~p\in \mathbb{N}^{*}\right)$, which showed the results $\left.\frac{v_y}{v_F}\right)_\text{max}=\left.\frac{v_y}{v_F}\right)_{q_2=\frac{1}{2}}$ and $\frac{1}{\pi}\leq \left.\frac{v_y}{v_F}\right)_\text{max} \leq \frac{3}{2\pi}$.

%%%%%%%%%%%%%%%%%%%%%%%%%%%%%%%
\section*{Acknowledgment}
%%%%%%%%%%%%%%%%%%%%%%%%%%%%%%

The generous support provided by the Saudi Center for Theoretical
Physics (SCTP) is highly appreciated by all authors.

%======================================================================
\section{Appendix}
%======================================================================
We explicitly determine the energy corresponding to different vertical contact points (VCPs) between bands. Indeed, in the Brillouin zone center $k_x = k_y = 0$ , we obtain
\begin{equation}\label{EqDispl}
  \varepsilon(k_x =k_y = 0)= q_1 \mathbb{V}_1+q_2 \mathbb{V}_2+q_3 \mathbb{V}_3+2m\pi,\qquad m\in \mathbb{Z}
\end{equation}
whereas in the Brillouin zone edge $\left(k_x =\pm \frac{\pi}{d},k_y = 0\right)$, the contact points should appear at energy
\begin{equation}\label{EqDisplpid}
  \varepsilon(k_x = \frac{\pi}{d},k_y = 0)=\varepsilon(k_x = -\frac{\pi}{d},k_y = 0)= q_1 \mathbb{V}_1+q_2 \mathbb{V}_2+q_3 \mathbb{V}_3+(2m\pm 1)\pi,\qquad m\in \mathbb{Z}.
\end{equation}
Both of \eqref{EqDispl} and \eqref{EqDisplpid} are valid for any distance $q_i\neq 1$ and potential $\mathbb{V}_i$. The pristine graphene submitted to the potential $\mathbb{V}_i$ $(q_i=1)$ has only one Dirac point located at $\varepsilon(k_x =k_y = 0)=\mathbb{V}_i$. Note that \eqref{EqDispl} generalizes that for SSLGSL-2R  \cite{Pham2010}.

For SSLGSL-3R, we look for the coordinates of the contact points  in the minibands $\left(k_x =\varepsilon = 0\right)$. This gives the relation
\begin{equation}
   \hspace*{-0.5em} \frac{\mathbb{V}^2-d^2 k_y^2 \cos \left((q_2-1) \sqrt{\mathbb{V}^2-d^2 k_y^2}\right)}{\mathbb{V}^2-d^2 k_y^2}\cosh (d k_y q_2)
  -\frac{d k_y  \sin \left((q_2-1) \sqrt{\mathbb{V}^2-d^2 k_y^2}\right)}{\sqrt{\mathbb{V}^2-d^2 k_y^2}}\sinh (d k_y q_2)=1
\end{equation}
which has different solutions according to the value of $q_2$. Indeed, for $q_2\neq 0$ we have only one solution $k_y=0$, but for $q_2=0$ (SSLGSL-2R) we have two solutions
\begin{equation}\label{ADP}
  k_y=0, \qquad k_{y_{D_m}}=\pm\frac{1}{d}\sqrt{\mathbb{V}^2-(2m\pi)^2}, \qquad \mathbb{V}^2>(2m\pi)^2
\end{equation}
with $m$ is an integer value no null. These show that for SSLGSL-2R, in minibands there are additional contact points (ACPs) ($k_x=0,k_y=k_{y_{D_m}},\varepsilon= 0$) in addition to ($k_x=0,k_y=0,\varepsilon= 0$). These results have been also obtained in  \cite{Pham2010}.

In order to determinate the dispersion relation close to a given contact point ($k_{x_c},k_{y_c},\varepsilon_c$), we expand \eqref{33} around it. To do this, we need to write \eqref{33} as an implicit function
\begin{equation}\label{implicit}
  f(k_x ,k_y,\varepsilon )=0.
\end{equation}
In the contact points, where the band structures are intersected, the gradient of dispersion relation \eqref{33} must be equal zero \cite{LIMA20151372}.
This implies that the contact point ($k_{x_c},k_{y_c},\varepsilon_c$) has to verify
\begin{equation}\label{grad}
  f(k_{x_c},k_{y_c},\varepsilon_c)=0, \qquad \bm{\nabla} f(k_{x_c},k_{y_c},\varepsilon_c)=0.
\end{equation}
Now, we consider the Taylor approximation of $f$ near the contact point ($k_{x_c},k_{y_c},\varepsilon_c$) to write
\begin{equation}\label{Taylor}
  f\left(k_x,k_y,\varepsilon\right)\approx f\left(k_{x_c},k_{y_c},\varepsilon_c\right)+\mathrm{\Delta}P\bm{\nabla} f\left(k_{x_c},k_{y_c},\varepsilon_c\right)+\frac{1}{2}\mathrm{\Delta}P^{t}
  \mathbb{H}f\left(k_{x_c},k_{y_c},\varepsilon_c\right)\mathrm{\Delta}P
\end{equation}
where $\mathrm{\Delta}P$ reads as
\begin{equation}
    \mathrm{\Delta}P=\left(\begin{array}{c}
            \mathrm{\Delta}k_x \\
            \mathrm{\Delta}k_y \\
            \mathrm{\Delta}\varepsilon
          \end{array}\right)=\left(\begin{array}{c}
            k_x-k_{x_c} \\
            k_y-k_{y_c} \\
            \varepsilon-\varepsilon_c
          \end{array}\right)
\end{equation}
and the Hessian matrix $\mathbb{H}$ is a square matrix of second-order partial derivatives of $f$, which can be explicitly determined by fixing the nature of contact point. Indeed, for SSLGSL-3R and in the contact point $(k_x=0,k_y=0,\varepsilon=2m\pi)$, the gradient and Hessian take the form
\begin{equation}\label{Hess111}
\bm{\nabla} f\left(0,0,2m\pi\right)=0, \qquad \mathbb{H}f\left(0,0,2m\pi\right)=\left(
  \begin{array}{ccc}
    -d^2 & 0 & 0 \\
    0 & B_1 & 0 \\
    0 & 0 & 1 \\
  \end{array}
\right)
\end{equation}
where the involved parameter is
\begin{eqnarray}\label{B1mNotZero}
  B_1(m\neq 0)&=&-\frac{d^2 \mathbb{V}^2}{8m^2\pi^2 \left(\mathbb{V}^2-\left(2m\pi\right)^2\right)^2} \left[12\pi^2 m^2 + \mathbb{V}^2 - 4\pi m (2\pi m+\mathbb{V})\cos(2\pi m q2+(1-q2)\mathbb{V})\right.\nonumber\\
  &&\left.+(2\pi m-\mathbb{V})((2\pi m+\mathbb{V})\cos(4\pi m q2)-4\pi m\cos(2\pi m q2+(q2-1)\mathbb{V}))\right]
\end{eqnarray}
which is only valid for $m \neq 0$ and $\mathbb{V} \neq 2\pi m$. However, for $m=0$ and $\mathbb{V} \neq 0$, we find a strictly negative quantity
\begin{equation}\label{B1mZero}
  B_1(m=0)=
  -\frac{d^2}{\mathbb{V}^2}\left[2+q_2^2\mathbb{V}^2-2\cos((q_2-1)\mathbb{V})-2q_2
  \mathbb{V}\sin((q_2-1)\mathbb{V})\right].
\end{equation}
In the Brillouin zone edge $\left(k_x=\pm\frac{\pi}{d},k_y=0,\varepsilon=\chi\right)$, the gradients are
\begin{equation}\label{grad777}
\bm{\nabla} f\left(\frac{\pi}{d},0,\chi\right)=
\bm{\nabla} f\left(-\frac{\pi}{d},0,\chi\right)=0
\end{equation}
and Hessians read as
\begin{equation}\label{Hess777}
\mathbb{H}f\left(\frac{\pi}{d},0,\chi\right)=
\mathbb{H}f\left(-\frac{\pi}{d},0,\chi\right)=\left(
  \begin{array}{ccc}
    d^2 & 0 & 0 \\
    0 & B_2\left(\chi\right) & 0 \\
    0 & 0 & -1 \\
  \end{array}
\right)
\end{equation}
where for $\mathbb{V}\neq |\chi|$ we have
\begin{eqnarray}\label{B2}
  B_2(\chi)&=&\frac{d^2\mathbb{V}^2}{2\chi^2(\chi^2-\mathbb{V}^2)^2}\Big[3\chi^2+
  \mathbb{V}^2+(\chi^2-\mathbb{V}^2)\cos(2\chi q_2)\nonumber\\
  &&+2\chi\Big(\left(\chi+\mathbb{V}\right)\cos\left(\mathbb{V}
  +\left(\chi-\mathbb{V}\right)q_2\right)+
  \left(\chi-\mathbb{V}\right)\cos\left(\mathbb{V}-(\chi+\mathbb{V})q_2\right)\Big)\Big]
\end{eqnarray}
which is strictly positive. For SSLGSL-2R in ACPs ($k_x=0,k_y=k_{y_{D_m}},\varepsilon= 0$), we find
\begin{equation}\label{Hess2}
\bm{\nabla} f\left(0,k_{y_{D_m}},0\right)=0, \qquad \mathbb{H}f\left(0,k_{y_{D_m}},0\right)=\left(
  \begin{array}{ccc}
    -d^2 & 0 & 0 \\
    0 & B_3 & 0 \\
    0 & 0 & \left(\frac{\mathbb{V}}{4m^2\pi^2}\right)^2 \\
  \end{array}
\right)
\end{equation}
and $B_3$ is giving by
\begin{equation}\label{B3}
  B_3=-d^2\left(\frac{\mathbb{V}^2-(2m\pi)^2}{4m^2\pi^2}\right)^2.
\end{equation}
All the above cases can be written in general form as
\begin{equation}\label{GHessian}
  \bm{\nabla} f\left(k_{x_c},k_{y_c},\varepsilon_c\right)=0, \qquad \mathbb{H}f\left(k_{x_c},k_{y_c},\varepsilon_c\right)=\left(
     \begin{array}{ccc}
       A & 0 & 0 \\
       0 & B & 0 \\
       0 & 0 & C \\
     \end{array}
   \right)
\end{equation}
where $A$, $B$ and $C$ can be fixed in terms of the parameters $(q, \mathbb{V}, d)$ for a given contact point. Using \eqref{implicit}, \eqref{grad} and \eqref{GHessian} to write \eqref{Taylor} in reduced form as
\begin{equation}\label{quadratic}
 A\mathrm{\Delta}k_x^2+B \mathrm{\Delta}k_y^2=-C \mathrm{\Delta}\varepsilon^2
\end{equation}
or equivalently
\begin{equation}\label{Diracdispersion}
  \varepsilon-\varepsilon_c=\pm \sqrt{-\frac{A}{C}\left(k_x-k_{x_c}\right)^2-\frac{B}{C}\left(k_y-k_{y_c}\right)^2}
\end{equation}
where sign($A$) = sign($B$) = $-$sign($C$). This is exactly the dispersion relation near the contact point and describes the anisotropy of energy contours.

On the other hand, from \eqref{Diracdispersion}, we can determine the two component of normalized group velocity corresponding to each contact point. They are defined by
\begin{equation}\label{GV}
  \frac{v_x}{v_F}=\frac{1}{d}\frac{\partial\varepsilon}{\partial k_x}, \qquad \frac{v_y}{v_F}=\frac{1}{d}\frac{\partial\varepsilon}{\partial k_y}.
\end{equation}
Near the contact points, we can approximate \eqref{GV} by using \eqref{quadratic} to obtain
\begin{equation}\label{GV3}
  \frac{v_x}{v_F}=\frac{1}{d}\left(\frac{\mathrm{\Delta}\varepsilon}{\mathrm{\Delta} k_x}\right)_{\mathrm{\Delta} k_y=0}=\frac{1}{d}\sqrt{-\frac{A}{C}}, \qquad \frac{v_y}{v_F}=\frac{1}{d}\left(\frac{\mathrm{\Delta}\varepsilon}{\mathrm{\Delta} k_y}\right)_{\mathrm{\Delta} k_x=0}=\frac{1}{d}\sqrt{-\frac{B}{C}}.
\end{equation}
Finally, using \eqref{GV3} to write \eqref{Diracdispersion} as
\begin{equation}\label{Diracdispersionyyy}
  \varepsilon-\varepsilon_c=\pm d \sqrt{\left(\frac{v_x}{v_F}\right)^2\left(k_x-k_{x_c}\right)^2+
  \left(\frac{v_y}{v_F}\right)^2\left(k_y-k_{y_c}\right)^2}
\end{equation}
which is exactly the Dirac-like form of the dispersion relation, derived also in \cite{Pham2014,Huy2014} and therefore the studied contact points can be seen as Dirac points. For Dirac fermions in pristine graphene $(v_x=v_y=v_F)$ and around the Dirac point $(k_x=0, k_y=0, \varepsilon= 0)$ \eqref{Diracdispersionyyy} reduces to
\begin{equation}\label{DiracdispersionPres}
  E=\frac{\hbar v_F}{d}\varepsilon=\pm \hbar v_F\sqrt{k_x^2+k_y^2}.
\end{equation}


\begin{thebibliography}{99}
 
 \bibitem{Novoselov666} K. S. Novoselov {\it et al.}, Science 306, 666 (2004).
 
  \bibitem{RN2} Y. Zhang, Y.-W. Tan, H. L. St\"ormer, and P. Kim, Nature 438, 201 (2005)
 
 \bibitem{RevModPhys.83.851} A. K. Geim, Rev. Mod. Phys. 83, 851 (2011).
 

 
 
 
 %\bibitem{PhysRevB.65.245420} Y. Zhang, Y.-W. Tan, H. L. Stormer, and P. Kim, Nature 438, 201 (2005).
 
 
 %\bibitem{PhysRevLett.95.146801} 
 \bibitem{PhysRevB.65.245420} Y. Zheng and T. Ando, Phys. Rev. B 65, 245420 (2002).
 
 %\bibitem{PhysRevB.76.075429} 
 
 \bibitem{PhysRevLett.95.146801}
 V. P. Gusynin and S. G. Sharapov, Phys. Rev. Lett. 95, 146801 (2005).
 
 \bibitem{2} N. Stander, B. Huard, and D. Goldhaber-Gordon, Phys. Rev. Lett. 102, 026807 (2009).
 
 \bibitem{3} M. I. Katsnelson, K. S. Novoselov, and A. K. Geim, Nature Phys. 2, 620 (2006).

 \bibitem{4} H. Sevincli, M. Topsakal, and S. Ciraci, Phys. Rev. B 78, 245402 (2008).
 
 \bibitem{CastroNeto2009} A. H. Castro Neto, F. Guinea, N. M. R. Peres, K. S. Novoselov, and A. K. Geim, Rev. Mod.
Phys. 81, 109 (2009).
 
 %\bibitem{PhysRevLett.100.056807} 
 \bibitem{PhysRevB.76.075429} 
 S. Marchini, S. Günther, and J. Wintterlin, Phys. Rev. B 76, 075429 (2007).

 %\bibitem{Jannik2008} 
 \bibitem{PhysRevLett.100.056807} 
 A. L. V\'{a}zquez de Parga {\it et al.}, Phys. Rev. Lett. 100, 056807 (2008).
 
 \bibitem{Jannik2008} J. C. Meyer, C. O. Girit, M. F. Crommie, and A. Zettl, Appl. Phys. Lett. 92, 123110 (2008).
 
 

 \bibitem{Park2008} C.-H. Park, L. Yang, Y.-W. Son, M. L. Cohen, and S. G. Louie, Nat. Phys. 4, 213 (2008).
 
 \bibitem{ParkCheol2008} C.-H. Park, L. Yang, Y.-W. Son, M. L. Cohen, and S. G. Louie, Phys. Rev. Lett. 101, 126804
(2008).

 \bibitem{ParkCheol2009} C.-H. Park, Y.-W. Son, L. Yang, M. L. Cohen, and S. G. Louie, Phys. Rev. Lett. 103, 046808
(2009).

 \bibitem{Brey2009} L. Brey and H. A. Fertig, Phys. Rev. Lett. 103, 046809 (2009).
 
 \bibitem{Barbier2010} M. Barbier, P. Vasilopoulos, and F. M. Peeters, Phys. Rev. B 81, 075438 (2010).
 
 \bibitem{Pham2010} C. H. Pham, H. C. Nguyen, and V. L. Nguyen, J. Phys. Condens. Matter 22, 425501 (2010).
 
 \bibitem{Ghosh2009} S. Ghosh and M. Sharma, J. Phys.: Condens. Matter 21, 292204 (2009).
 
 \bibitem{Ramezani2010} M. R. Masir, P. Vasilopoulos, and F. M. Peeters, J. Phys.: Condens. Matter 22, 465302 (2010).
 \bibitem{Snyman2009} I. Snyman, Phys. Rev. B 80, 054303 (2009).
 
\bibitem{DellAnna2011}  L. Dell'Anna and A. De Martino, Phys. Rev. B 83, 155449 (2011).

 \bibitem{VQui2012} V. Q. Le, C. H. Pham, and V. L. Nguyen, J. Phys.: Condens. Matter 24, 345502 (2012).
 
 \bibitem{LIMA20151372} J. R. Lima, Phys. Lett. A 379, 1372 (2015).
 
 \bibitem{Huy2014} C. H. Pham and V. L. Nguyen, J. Phys.: Condens. Matter 26, 425502 (2014).
 
 \bibitem{Pham2014} C. H. Pham, T. T. Nguyen, and V. L. Nguyen, J. Appl. Phys. 116, 123707 (2014).
 
\end{thebibliography}
\end{document}